\documentclass[twocolumn,numberedappendix,breaklinks=true]{aastex62}

\usepackage{natbib}
\usepackage{multirow}
\usepackage{gensymb} 
\usepackage[flushleft]{threeparttable} 
\usepackage{amsmath}
\usepackage{url}
\usepackage{multirow}
\usepackage{wasysym}

\setcitestyle{comma}

\shorttitle{SMF of galaxies at $z\sim6$ to $\sim10$ from GREATS}
\shortauthors{Stefanon et al.}

\begin{document}

\title{Galaxy Stellar Mass Functions from $z\sim 10$ to $z\sim 6$ using the Deepest Spitzer/IRAC Data:  No Significant Evolution in the Stellar-to-Halo Mass Ratio of Galaxies in the First Gyr of Cosmic Time}

\author{Mauro Stefanon}
\affiliation{Leiden Observatory, Leiden University, NL-2300 RA Leiden, Netherlands}

\author{Rychard J. Bouwens}
\affiliation{Leiden Observatory, Leiden University, NL-2300 RA Leiden, Netherlands}

\author{Ivo Labb\'e}
\affiliation{Centre for Astrophysics and SuperComputing, Swinburne, University of Technology, Hawthorn, Victoria, 3122, Australia}

\author{Garth D. Illingworth}
\affiliation{UCO/Lick Observatory, University of California, Santa Cruz, 1156 High St, Santa Cruz, CA 95064, USA}

\author{Valentino Gonzalez}
\affiliation{Departamento de Astronom\'ia, Universidad de Chile, Casilla 36-D, Santiago 7591245, Chile}
\affiliation{Centro de Astrof\'isica y Tecnologias Afines (CATA), Camino del Observatorio 1515, Las Condes, Santiago 7591245, Chile}

\author{Pascal A. Oesch}
\affiliation{Departement d'Astronomie, Universit\'e de Gen\'eve, 51 Ch. des Maillettes, CH-1290 Versoix, Switzerland}
\affiliation{International Associate, Cosmic Dawn Center (DAWN), Niels Bohr Institute, University of Copenhagen and DTU-Space, Technical University of Denmark}

\email{Email: stefanon@strw.leidenuniv.nl}

\begin{abstract}

We present new stellar mass functions at $z\sim6$, $z\sim7$, $z\sim8$, $z\sim9$ and, for the first time, $z\sim10$, constructed from $\sim800$ Lyman-Break galaxies previously identified over the XDF/UDF, parallels and the five CANDELS fields. Our study is distinctive due to (1) the much deeper ($\sim200$~hour) wide-area \textit{Spitzer}/IRAC imaging at $3.6\mu$m and $4.5\mu$m from the GOODS Re-ionization Era wide Area Treasury from Spitzer (GREATS) program and (2) consideration of $z\sim6-10$ sources over a $3\times$ larger area than previous \textit{HST}+\textit{Spitzer} studies. The \textit{Spitzer}/IRAC data enable $\ge2\sigma$ rest-frame optical detections for an unprecedented $50\%$ of galaxies down to a stellar mass limit of $\sim10^{8}\mathcal{M}_\odot$ across all redshifts. Schechter fits to our volume densities suggest a combined evolution in characteristic mass $\mathcal{M}^*$ and normalization factor $\phi^*$ between $z\sim6$ and $z\sim8$. The stellar mass density (SMD) increases by $\sim1000\times$ in the $\sim500$\,Myr between $z\sim10$ and $z\sim6$, with indications of a steeper evolution between $z\sim10$ and $z\sim8$, similar to the previously-reported trend of the star-formation rate density. Strikingly, abundance matching to the Bolshoi-Planck simulation indicates halo mass densities evolving at approximately the same rate as the SMD between $z\sim10$ and $z\sim4$. Our results show that the stellar-to-halo mass ratios, a proxy for the star-formation efficiency, do not change significantly over the huge stellar mass build-up occurred from $z\sim10$ to $z\sim6$, indicating that the assembly of stellar mass closely mirrors the build-up in halo mass in the first $\sim1$\,Gyr of cosmic history. JWST is poised to extend these results into the "first galaxy" epoch at $z\gtrsim10$.
\end{abstract}


\section{Introduction}

In the last decade, the increased sensitivity at near-infrared wavelengths provided by the \textit{Hubble Space Telescope (HST)} Wide Field Camera 3 (WFC3) has revealed $\gtrsim 10$k galaxies at $z\gtrsim4$ (e.g., \citealt{bouwens2015, finkelstein2015a}), probing galaxy formation to epochs as early as $z\sim10-12$, just $\sim400-500$\,Myr after the Big Bang (see e.g., \citealt{bouwens2011,bouwens2013, ellis2013,coe2013,mclure2013,oesch2014,oesch2016,oesch2018, mcleod2016, calvi2016,salmon2018, morishita2018,lam2019b}). 

Despite the remarkable advances in the field, some uncertainties still exist on the estimates of fundamental parameters such as the cosmic star-formation rate density (CSFRD). A number of studies suggest that the CSFRD underwent a rapid increase in the first $\sim600$\,Myr, followed by a less rapid growth (see e.g., \citealt{oesch2012, oesch2014, oesch2018, ellis2013, bouwens2015}), consistent with the rate of growth of the dark matter halos (e.g., \citealt{oesch2018}). Other works, however, indicate higher densities at $z\sim8-10$ resulting in a reduced evolution of the CSFRD  from $z\sim9$ to $z\sim4$ (e.g., \citealt{mclure2013, mcleod2016, bhatawdekar2019,kikuchihara2020}). Furthermore, the discovery of enigmatic objects such as GN-z11  (\citealt{oesch2016}) and MACS1149-JD1 (\citealt{zheng2012, hoag2018, hashimoto2018}) prompt questions about how such massive galaxies could assemble so rapidly.

A complementary approach to studying the assembly of galaxies consists of measuring the integral with cosmic time of the SFR,  i.e., the stellar mass ($\mathcal{M}_\star$). Numerous studies have estimated the stellar mass function (SMF) and the stellar mass density (SMD) of galaxies to $z\sim8$ (see \citealt{madau2014} and references therein, and those we list in Sect. \ref{sect:SMD}). These two approaches should yield consistent results. The emerging picture is that from $z\sim7$ to today the evolution of the SMD  is actually consistent with that expected from the integration of the CSFRD (modulo a systematic offset of $\sim0.2-0.5$\,dex - see e.g, \citealt{madau2014}, \citealt{leja2019} and references therein). 

At higher redshifts, the current estimates suggest a marginal evolution of the SMD for $8\lesssim z\lesssim9$ (e.g., \citealt{bhatawdekar2019, kikuchihara2020}) followed by a $\gtrsim1-1.5$\,dex drop by $z\sim10$ (\citealt{oesch2014}). Nevertheless, current SMF measurements at such high redshifts suffer from higher systematic uncertainties, both at the sample selection stage and in the estimates of stellar masses for individual sources, as we discuss below.

A number of recent papers have estimated SMFs from samples selected over areas ranging from $\sim$few$\times$arcmin$^2$ to $\sim100$\,arcmin$^2$ (e.g., \citealt{duncan2014, grazian2015, song2016, bhatawdekar2019, kikuchihara2020}). Such small areas, however, introduce large cosmic variance, particularly at the massive end, with uncertanties from cosmic variance approaching  $\sim50\%$ at $z\sim7-8$ (see e.g., \citealt{mcleod2020} and \citealt{bhowmick2020}), while the corresponding small sample sizes for massive galaxies result in larger Poissonian uncertainties. The obvious solution, observations over $\sim$square-degree fields, provide stringent constraints on the high-mass end, but lack sufficient depth to constrain the low-mass end (e.g., \citealt{davidzon2017}). The lack of deep wide-field areas has been a challenge for deriving robust SMFs.

\begin{figure}
\includegraphics[width=9cm]{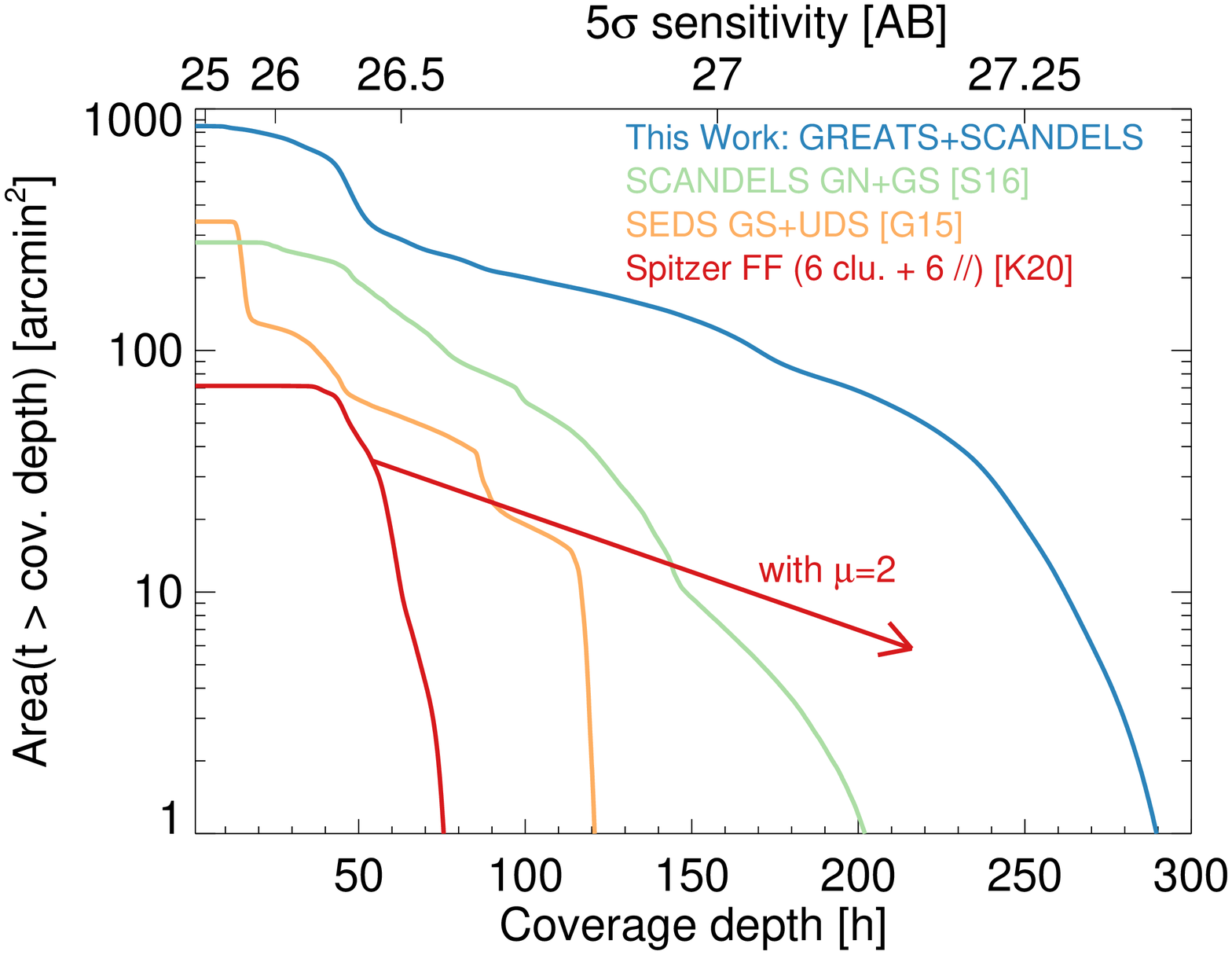}
\caption{Cumulative area as a function of coverage depth (in hr) in the IRAC $3.6\mu$m band, for representative sets of observations recently adopted for the measurement of the SMF at $z\sim6$ and above. Specifically, we include data from GREATS (\citealt{stefanon2021a}), S-CANDELS (\citealt{ashby2015}), SEDS (\citealt{ashby2013}) and the Spitzer-Frontier Fields (\citealt{shipley2018}). The top axis presents approximate point-source $5\sigma$ sensitivity from the SENS-PET calculator. The red arrow marks the effective depth that can be obtained when sources in the cluster fields are magnified by $\mu=2$. The smaller area results from the decrease of effective area due to magnification (a factor $\sim6$ for $z\sim8$ sources when $\mu\ge2$). In square brackets we indicate the studies with their adopted set of IRAC data: G15: \citet[see also \citealt{duncan2014}]{grazian2015}; S16: \citet{song2016}; K20: \citet[see also \citealt{bhatawdekar2019}]{kikuchihara2020}. The very substantial gains from the new GREATS + S-CANDELS datsets are apparent.\label{fig:irac}}
\end{figure}

Robust estimates of stellar masses require detections in the  rest-frame optical.  At redshifts $z\gtrsim5$  these can currently only be obtained by observations with \textit{Spitzer}/IRAC (\citealt{fazio2004}). The current depths of IRAC data in extragalactic fields allow for individual detections of only the brightest, and hence most massive sources (see Figure \ref{fig:irac}). Consequently, spectral energy distributions (SEDs) in the optical at lower masses are still lacking, or at best quite uncertain, esentially restricting the derivation of the SMF to relatively massive systems at high redshifts (e.g., \citealt{duncan2014, grazian2015, song2016}). Furthermore, the use of stacking to characterize the properties of fainter sources has only resulted in modest gains due to the small sample sizes (e.g., \citealt{gonzalez2012, song2016, kikuchihara2020}). An interesting exception to this limitation are recent studies based on the Hubble Frontier Field (HFF - \citealt{lotz2017}) initiative, which leverage the gravitational magnifications of low-z galaxy clusters to reach fainter limits at high redshifts (e.g., \citealt{bhatawdekar2019, kikuchihara2020}). Unfortunately,  systematic uncertainties in the magnification maps (e.g., \citealt{bouwens2017}) and the higher surface densities of nearby large and bright objects in these fields (e.g. \citealt{merlin2016b,castellano2016, shipley2018}) make it very difficult to carry out reliable photometry.

New IRAC data combined with \textit{Hubble} observations now provide an opportunity to overcome the aforementioned challenges. In this work, we measure the galaxy SMF at $z\sim6-10$ using the most comprehensive selection of $z\sim6-10$ galaxies from the \textit{HST} legacy fields, including galaxies from all five CANDELS fields (\citealt{grogin2011, koekemoer2011}). Most importantly, new full-depth IRAC mosaics from the GOODS Re-ionization Era wide-Area Treasury from Spitzer (GREATS - PI: I. Labb\'e, \citealt{stefanon2021a}) allow us to determine their rest-frame optical fluxes. These data provide $\ge 2\sigma$ detections in the IRAC $3.6\mu$m and $4.5\mu$m bands for $50\%$ of individual sources in the sample down to stellar masses $\mathcal{M}_\star\sim10^8\mathcal{M}_\odot$ over most of the considered redshift range.   Additionally, our galaxy SMFs leverage a search area that is $3\times$ larger than previous studies, lessening both the impact of cosmic variance and Poisson noise (by $1.7\times$). The combination of the new GREATS dataset and the large Hubble sample enables derivation of an SMF where sample statistics and cosmic variance are minimized, as well as providing the needed rest-frame optical SEDs for more accurate mass estimates.

\begin{figure*}
\includegraphics[width=18.5cm]{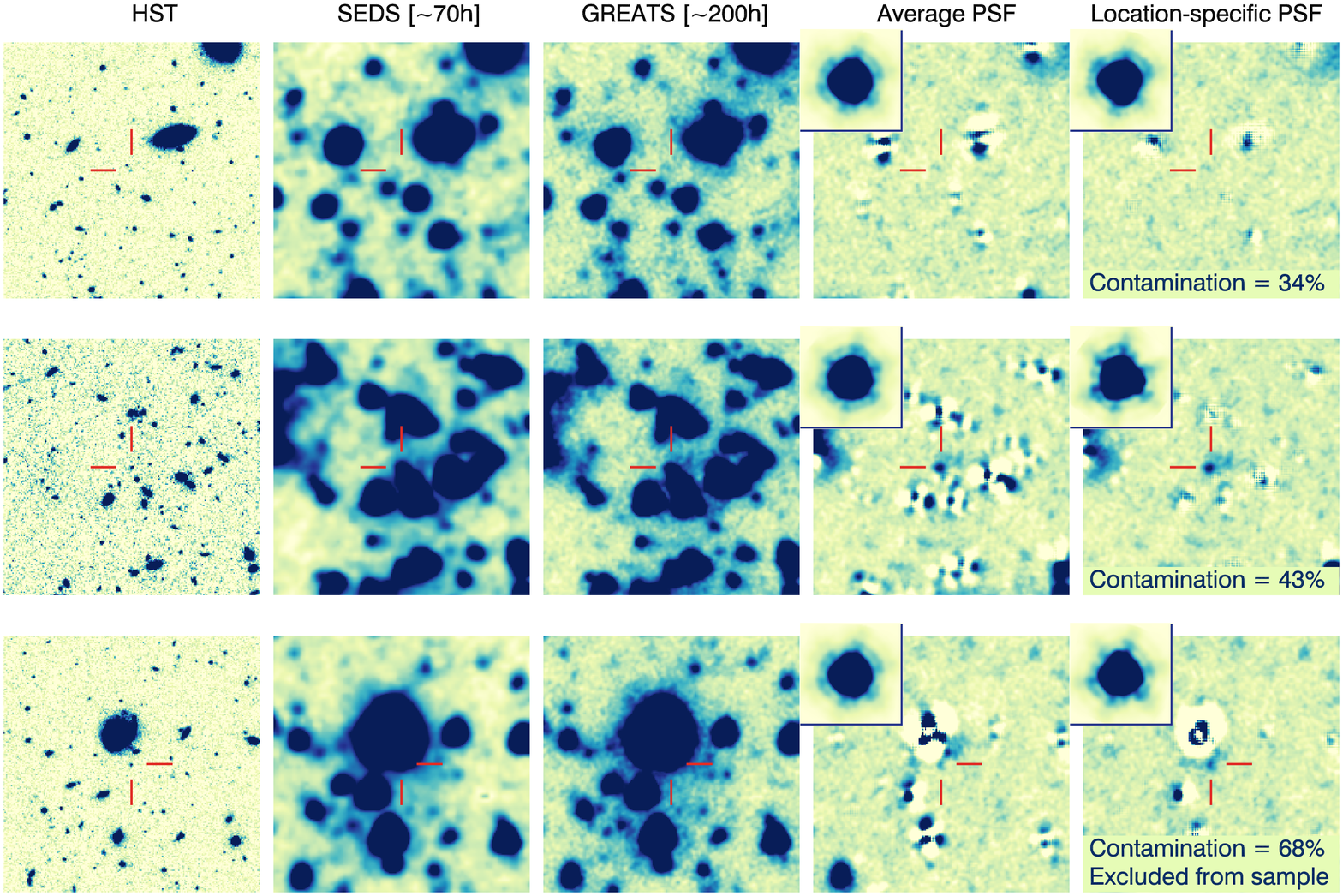}
\caption{Illustration of our sophisticated procedures for handling the deep IRAC data used in this work. Each row refers to a specific object in the $z\sim7-8$ compilation of \citet{bouwens2015} which constitute our initial sample (top-to-bottom are GSDZ-2460945596, GNWZ-7268117400  and GSDZ-2288549126). Each stamp is $\sim30''$ per side and, in each row, they match to the same region of sky. In each stamp, the location of the high-z source corresponds to the intersection of the two red segments. Left to right, columns present the combined image from HST ($J_{125}+JH_{140}+H_{160}$), an image stamp at $3.6\mu$m from SEDS \citep{ashby2013}, which corresponds to a nominal coverage depth of $\sim70$\,hrs (including also the GOODS IRAC data), and the same region in the $200$\,hrs GREATS $3.6\mu$m mosaic. In the last two columns we present our results subtracting neighbouring sources with \textsc{Mophongo}  adopting, first, an average PSF and, second, the specific PSF reconstructed at the location of the source, accounting for the orientations of all contributing observations (as first pioneered in \citealt{labbe2015}), respectively. The estimated contamination from neighbouring sources is reported at the bottom of the right-most panel. All IRAC stamps share the same flux density cuts. The adopted PSF is shown in the top-left corner of the corresponding stamp. As can clearly be seen in the rightmost column, the combination of increased depth and accurate PSF reconstruction from the location-specific PSF allow us to obtain more robust flux densities in the IRAC bands. The last row shows an object excluded from our final sample because of the large contamination ($>65\%$) from the very bright neighbouring source which made the photometry more uncertain, even though the removal of the neighbours was reasonably successful.\label{fig:example}}
\end{figure*}

A brief summary of the organization of this paper follows.   In  Section \ref{sect:sample}, we briefly describe the sample adopted for the SMF measurements.   Section \ref{sect:mstar_0} details the procedures we followed to estimate the stellar mass of galaxies depending on the redshift bin and on the significance of the IRAC detections.  In Section \ref{sect:completeness} we characterize the completeness of our sample. Section \ref{sect:results} includes a presentation of our new SMF determinations and compares these new results with others in the literature. In Section \ref{sect:discussion} we characterize the build-up of the SMD with cosmic time and connect our results to a similar build-up in the dark matter halo mass density and limited evolution in the stellar-to-halo mass ratios.  In Section  \ref{sect:conclusions}, we include a summary.

Throughout this paper we adopt magnitudes in the AB system (\citealt{oke1983}), and a $\Lambda$CDM cosmology with $\Omega_\mathrm{m}=0.3$, $\Omega_\Lambda=0.7$, $H_0=70$\,km/s/Mpc, unless otherwise stated. Our stellar mass measurements assumed a \citet{salpeter1955} initial mass function (IMF). We conventionally denote  the logarithm in base $10$ with $\log$.

\section{Samples}
\label{sect:sample}

\begin{deluxetable*}{llccccc}
\tablecaption{Observational data used for the SMF estimates. \label{tab:obs_data}}
\tablehead{\multicolumn{2}{c}{Field} & \colhead{Area} & \colhead{$H_\mathrm{160}$\tablenotemark{a}} &\colhead{IRAC Data\tablenotemark{b}} & \colhead{$3.6\mu$m\tablenotemark{c}} & \colhead{$4.5\mu$m\tablenotemark{c}}    \\
\multicolumn{2}{c}{Name} & \colhead{[arcmin$^2$]} & \colhead{$5\sigma$ [mag]} & & \colhead{$5\sigma$ [mag]} & \colhead{$5\sigma$ [mag]}  
}
\startdata
\multicolumn{2}{l}{XDF}      & $4.7$   & $29.4$ & GREATS        & $\sim27.2$  & $\sim26.7$   \\
\multicolumn{2}{l}{HUDF09-1} & $4.7$   & $28.3$ & GREATS        & $\sim26.3$  & $\sim25.8$   \\
\multicolumn{2}{l}{HUDF09-2} & $4.7$   & $28.7$ & GREATS        & $\sim27.0$  & $25.5-26.0$  \\
\multicolumn{2}{l}{ERS}      & $40.5$  & $27.4$ & GREATS        & $26.2-27.0$ & $25.6-26.7$  \\
CANDELS & GOODS-N Deep       & $62.9$  & $27.5$ & GREATS        & $27.0-27.3$ & $26.5-26.8$  \\
        & GOODS-N Wide       & $60.9$  & $26.7$ & GREATS        & $26.3-27.2$ & $25.8-26.8$  \\
        & GOODS-S Deep       & $64.5$  & $27.5$ & GREATS        & $\sim27.3$  & $26.6-26.9$  \\
        & GOODS-S Wide       & $34.2$  & $26.8$ & GREATS        & $26.5-27.2$ & $26.2-26.7$  \\
        & COSMOS             & $151.9$ & $26.8$ & SEDS+S-CANDELS & $26.4-26.7$ & $26.0-26.3$  \\
        & EGS                & $150.7$ & $26.9$ & SEDS+S-CANDELS & $26.1-26.5$ & $25.7-26.1$  \\
        & UDS                & $151.2$ & $26.8$ & SEDS+S-CANDELS & $25.4-26.3$ & $25.0-25.9$  \\[5pt]
\hline
\multicolumn{2}{c}{Totals:} & $730.9$ & & & &  \\
\enddata
\tablenotetext{a}{$5\sigma$ limit from \citet{bouwens2015}, computed from the median of measured uncertainties of sources.}
\tablenotetext{b}{GREATS: \citet{stefanon2021a}; SEDS: \citet{ashby2013a}; S-CANDELS: \citet{ashby2015}.}
\tablenotetext{c}{Nominal $5\sigma$ limit for point sources from the SENS-PET exposure time calculator, based on the exposure time maps. Due to inhomogeneities in the coverage, a range of values is quoted when the depth varies by more than $\sim0.2$\,mag across the field. Because of the combined effects of the broad \textit{Spitzer}/IRAC PSF and the long exposure times, source blending may reduce the actual depth (see discussion in \citealt{labbe2015}).}
\end{deluxetable*}

\begin{deluxetable*}{llccccc}
\tablecaption{Number of sources in the samples used for our SMF measurements.  \label{tab:sample}}
\tablehead{\multicolumn{2}{c}{Field} & \multicolumn{5}{c}{\# Sources\tablenotemark{a}}   \\
\multicolumn{2}{c}{Name} & \colhead{$z\sim6$} &\colhead{$z\sim7$} &\colhead{$z\sim8$} &\colhead{$z\sim9$} &\colhead{$z\sim10$}   
}
\startdata
\multicolumn{2}{l}{XDF}      & $ 30$ $( 17)$ & $  7$ $(  3)$ & $  8$ $(  5)$ & $6$ $(1)$ & $2$ $(0)$  \\
\multicolumn{2}{l}{HUDF09-1} & $ 15$ $(  4)$ & $  7$ $(  1)$ & $  3$ $(  0)$ & $0$       & $0$        \\
\multicolumn{2}{l}{HUDF09-2} & $ 11$ $(  7)$ & $  6$ $(  2)$ & $  2$ $(  1)$ & $1$ $(1)$ & $0$        \\
\multicolumn{2}{l}{ERS}      & $ 38$ $( 30)$ & $ 15$ $( 14)$ & $  2$ $(  0)$ & $1$ $(1)$ & $0$        \\
CANDELS & GOODS-N Deep       & $ 89$ $( 73)$ & $ 70$ $( 47)$ & $ 14$ $(  5)$ & $2$ $(2)$ & $2$ $(1)$  \\
        & GOODS-N Wide       & $ 51$ $( 41)$ & $ 24$ $( 19)$ & $ 10$ $(  6)$ & $0$       & $1$ $(1)$  \\
        & GOODS-S Deep       & $114$ $( 90)$ & $ 37$ $( 23)$ & $ 15$ $( 11)$ & $1$ $(1)$ & $1$ $(1)$  \\
        & GOODS-S Wide       & $ 36$ $( 31)$ & $  6$ $(  5)$ & $  0$         & $1$ $(1)$ & $0$        \\
        & COSMOS             & $ 37$ $( 33)$ & $ 15$ $( 12)$ & $  5$ $(  5)$ & $1$ $(1)$ & $0$        \\
        & EGS                & $ 71$ $( 62)$ & $\ldots$\tablenotemark{$\dagger$}      & $\ldots$\tablenotemark{$\dagger$}      & $5$ $(5)$ & $0$        \\
        & UDS                & $ 31$ $( 28)$ & $ 17$ $( 16)$ & $  6$ $(  5)$ & $1$ $(1)$ & $0$        \\[5pt]
\hline
\multicolumn{2}{c}{Totals:} & $523$ $(416)$ & $204$ $(142)$ & $65$ $(38)$ & $19$ $(14)$ & $6$ $(3)$ \\
\enddata
\tablenotetext{a}{Number of sources selected in each redshift bin. The quantities in parentheses indicate the number of sources with S/N$>2$ in the IRAC bands.}
\tablenotetext{\dagger}{We excluded the $z\sim7$ and $z\sim8$ samples in EGS because of \citet{bouwens2015}'s use of the IRAC data itself (given the lack of deep $Y$-band data for this field) to help with the selection of these sources and thus large uncertainties on the photometric redshifts of $z=7-8$ sources from the EGS field.}
\end{deluxetable*}

For this study we set out to derive the SMF in redshift bins centered at $z\sim6, 7, 8, 9$ and $10$.  In the next sections we outline our sample selection criteria, while in Tables \ref{tab:obs_data} and \ref{tab:sample} we summarize the main properties of the adopted datasets and of the resulting samples.

\subsection{Samples at $z\sim6, 7$ and $8$}

Our goal was to make use of the largest and most comprehensive set of $z\sim6, 7$ and $8$ galaxies from the CANDELS fields and assorted deep \textit{HST} fields for the purposes of deriving galaxy stellar mass functions. Specifically, the $z\sim6, 7$ and $8$ samples we utilize are based on  the $I-$, $z-$ and $Y-$dropouts, respectively, \citet{bouwens2015} identified over the CANDELS (\citealt{grogin2011, koekemoer2011}) GOODS-N, GOODS-S (\citealt{giavalisco2004}), UDS (\citealt{lawrence2007}) and COSMOS (\citealt{scoville2007}) fields, the ERS field (\citealt{windhorst2011}), and the UDF/XDF (\citealt{beckwith2006,illingworth2013, ellis2013}) with the HUDF09-1 and HUFD09-2 parallels (\citealt{bouwens2011b}). We also included the $z\sim6$ candidates \citet{bouwens2015} identified over the CANDELS EGS field (\citealt{davis2007}), but not the $z\sim7-8$ candidates from this field given the lack of deep $Y$-band imaging to segregate galaxies at $z\sim7$ from those at $z\sim8$.

The CANDELS fields have received substantial coverage with the \textit{Spitzer} Infrared Array Camera (IRAC - \citealt{fazio2004}), in particular at $3.6\mu$m and $4.5\mu$m.  Starting at $z\sim5$, these bands probe the rest-frame optical, pivotal for the estimates of stellar masses. Furthermore, the evolution with redshift of the $[3.6]-[4.5]$ color suggests contributions by strong emission lines such as [\ion{O}{2}]$\lambda3727$, [\ion{O}{3}]$\lambda\lambda 4959,5007$, H$\alpha$ and $H\beta$. Inclusion of these lines into the fitting process can greatly improve the accuracy of the photometric redshifts (e.g. \citealt{smit2014, roberts-borsani2016}) and better discriminate against lower redshift interlopers.

Most importantly, and a crucial addition to the goals of this study, the GOODS-N and GOODS-S fields benefit from new full-depth  \textit{Spitzer}/IRAC $3.6\mu$m and $4.5\mu$m imaging from the \textit{GOODS Re-ionization Era wide-Area Treasury from Spitzer} (GREATS) program (PI: I. Labb\'e, \citealt{stefanon2021a}). GREATS increases the integration time to $\gtrsim200$\,hr over an area of $\sim100$\,arcmin$^2$, while improving the homogeneity in both $3.6\mu$m and $4.5\mu$m depths.  The corresponding IRAC point-source $5\sigma$ sensitivity of $\sim27.2$\,mag approximately matches the HST $H_{160}$ flux density limits from CANDELS. 

For the EGS, UDS and COSMOS fields we included observations from the S-CANDELS program (\citealt{ashby2015}), which, in combination with the SEDS program (\citealt{ashby2013a}), provides a coverage of $\gtrsim50$\, hr per field (nominal SENS-PET\footnote{\url{http://ssc.spitzer.caltech.edu/warmmission/propkit/pet/senspet/}} $5\sigma$ limits for point sources of $\sim26.0-26.4$\,mag at $3.6\mu$m and $4.5\mu$m, respectively).

Figure \ref{fig:irac} presents the cumulative area as a function of integration time for the mosaics adopted in our study,  and for few other prior IRAC datasets that have been used in recent SMF determinations at $z>6$ (\citealt{duncan2014, grazian2015, song2016, bhatawdekar2019} and \citealt{kikuchihara2020}). Our data are $\gtrsim2\times$ deeper over the GOODS fields and reach $\sim3\times$ more area thanks to the combination of all the CANDELS fields. The IRAC data adopted for our study provide moderate to high S/N information for a large fraction of sources in our sample (we further discuss this in Section \ref{sect:completeness}).

We extracted new flux densities from the GREATS and S-CANDELS mosaics  for all sources in our sample using the deblending code \textsc{Mophongo} (\citealt{labbe2006, labbe2010a, labbe2010b, labbe2013, labbe2015}).  In Figure \ref{fig:example} we present image stamps of three $z\sim7-8$ sources as they appear in the $\sim70$\,hr-deep IRAC mosaics from SEDS and in the $\sim200$\,hr regions of  GREATS. In the same figure we also show the residuals after subtracting their neighbours with \textsc{Mophongo} adopting first an average PSF and then second the PSF reconstructed accounting for the specific orientations of the IRAC observations over the corresponding regions (as pioneered in earlier work by \citealt{labbe2015}). It is evident from the last two columns of Figure \ref{fig:example} how, not only the photometric depth, but also an accurate knowledge of the PSF, are of crucial importance for a robust flux density estimate using Spitzer data.

\begin{figure*}
\includegraphics[width=19cm]{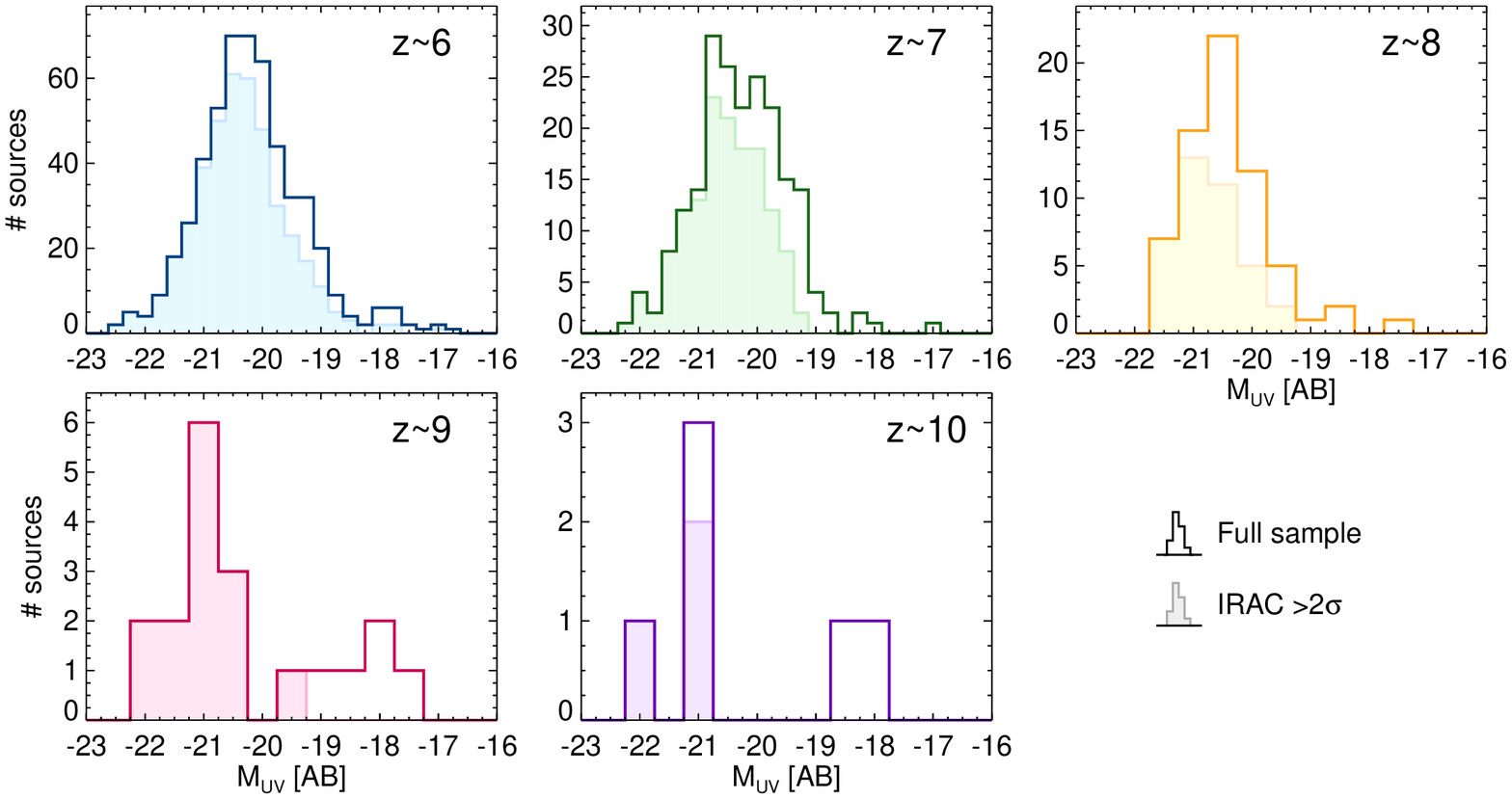}
\caption{UV luminosity distribution  of the sources in our samples after removing objects with potentially high contamination from neighbours in each of the IRAC bands. The corresponding redshift bin for each panel is shown in the top-right corner. In each panel, the histogram marked by the darker line corresponds to the full sample, while the filled histogram corresponds to those sources with S/N$>2$ in both the IRAC bands ($z\sim6,7$ and $8$) or in the $4.5\mu$m band only ($z\sim9$ and $10$).  \label{fig:sample}}
\end{figure*}

We redetermined the photometric redshifts of our sample with \textsc{EAzY} \citep{brammer2008}, complementing the standard template set with templates extracted from the Binary Population and Spectral Synthesis code (BPASS - \citealt{eldridge2017}) v1.1 for metallicity  $Z = 0.2Z_\odot$. We incorporated nebular lines with equivalent widths EW$($H$\alpha)\sim 1000-3000$\AA\ and line ratios from \citet{anders2003}, as these extreme EW reproduce the observed $[3.6] -[4.5]$ colors for many spectroscopically confirmed $z \sim 7-9$ galaxies (\citealt{ono2012, finkelstein2013, oesch2015b, roberts-borsani2016, zitrin2015, stark2016}). Driven by recent observational results (e.g., \citealt{roberts-borsani2016, oesch2015b, zitrin2015, stark2017, debarros2017}), we removed the Ly$\alpha$ line from those templates that had EW(Ly$\alpha$)$>40$\,\AA. We also included templates of $2$\, Gyr-old, passively evolving systems from \citet{bruzual2003}, with \citet{calzetti2000} extinction in the range $A_V = 0 - 8$\, mag to test the robustness of our selected candidates against being lower-redshift interlopers that were highly attenuated by dust. To further improve the robustness of the $z\sim6, 7$ and $z\sim8$ samples, we required the integral of the redshift likelihood (equivalent to a posterior probability assuming a uniform prior) to be $p(z)>0.6$ beyond $z=5, 6,$ and $7$, respectively, and the peak of the $p(z)$ to lie within the ranges $5.5\le z_\mathrm{phot}<6.3$, $6.3\le z_\mathrm{phot}<7.5$ and $7.5\le z_\mathrm{phot}<8.5$ for the samples at $z\sim6, 7$ and $z\sim8$, respectively. These constraints had a modest impact on the final sizes of our samples removing $22\pm5\%$ and $25\pm10\%$ ($p(z)$ and $z_\mathrm{phot}$ selections, respectively).  After applying this criteria our samples included $789, 357$ and $131$ sources, respectively.

Finally, to reduce potential systematics in the stellar mass estimates, we removed  from our sample any sources with $\geq 65\%$ flux contamination from neighbouring objects\footnote{We define the contamination $c$ to be $c=\sum f_\mathrm{n}/(f_\mathrm{s}+\sum f_\mathrm{n})$, where $f_\mathrm{s}$ is the flux density estimated for the source in a $1\farcs8$-diameter aperture, and $\sum f_\mathrm{n}$ the cumulative flux density from all neighbouring sources entering that aperture.}  in either IRAC band. After this step the sample included $523, 204$, and $65$ objects at $z\sim6,7$ and $8$ (corresponding to $\sim66\%, 57\%$ and $50\%$ of the parent sample), respectively. In Figure \ref{fig:sample}, we present our final sample in terms of UV luminosity, with sources segregated by the significance of the associated IRAC measurements. This Figure and Table \ref{tab:sample} indicate that, for the $z\sim6-8$ samples, $\sim25\%$ of sources remain undetected (at $2\sigma$) in at least one of the IRAC bands. To account for this selection in our SMF estimates, we implemented the Monte Carlo simulation described in Appendix \ref{app:LF_cc}. The estimated statistical corrections allow us to recover the UV LF over the full range of absolute magnitudes, indicating that we can confidently measure the corresponding SMFs (see Figure \ref{fig:LF_cc} of the Appendix). However, the median of the corrections become very large ($>10\times$) for  $M_\mathrm{UV}\gtrsim -16.75, -17.25$ and $\sim -17.5$\,mag at $z\sim6, 7$ and $\sim8$, respectively, making the associated volume densities more uncertain. For this reason, in our analysis we flag those measurements that are  affected by very large corrections.

\subsection{Sample at $z\sim9$}

The initial $z\sim9$ sample included the $YJ$-dropouts from \citet{oesch2014, bouwens2016} and \citet{bouwens2019a} identified over the five CANDELS fields, and are summarized in Table~2 of \citet{bouwens2019a}. For consistency with the $z\sim6-8$ selection criteria, we excluded GS-z9-5 and UDS910-5 because their probability of being genuine $z>8$ sources, $p(z>8)\sim0.55$ and $0.58$, respectively, does not satisfy our threshold ($p(z>8)=0.6$). We complemented this sample with GN-z10-3 from \citet{oesch2014}, which has a photometric redshift of $z_\mathrm{phot}=9.5$,  and $6$  sources identified by \citet{oesch2013} over the XDF region (we excluded XDFyj-39446317 due to uncertainties on its high-z nature - see \citealt{oesch2013} for details). This resulted in a total of $19$ sources. Given the availability of updated, deeper IRAC $3.6\mu$m and $4.5\mu$m mosaics from GREATS, we measured new flux densities in those bands for all sources in the GOODS fields using the same procedures described in the previous Section.

\subsection{Sample at $z\sim10$}
 For the $z\sim10$ sample we adopted the compilation of \citet{oesch2018} which includes sources identified over the GOODS-N, GOODS-S and XDF fields. We complemented this sample with one additional $J-$dropout identified in the XDF field by \citet[XDFJ-4023680031]{bouwens2015}, for a total of $6$ sources. For all sources we measured new $3.6\mu$m and $4.5\mu$m flux densities from GREATS using the same methods described in Section \ref{sect:sample}.

\begin{figure*}
\includegraphics[width=18cm]{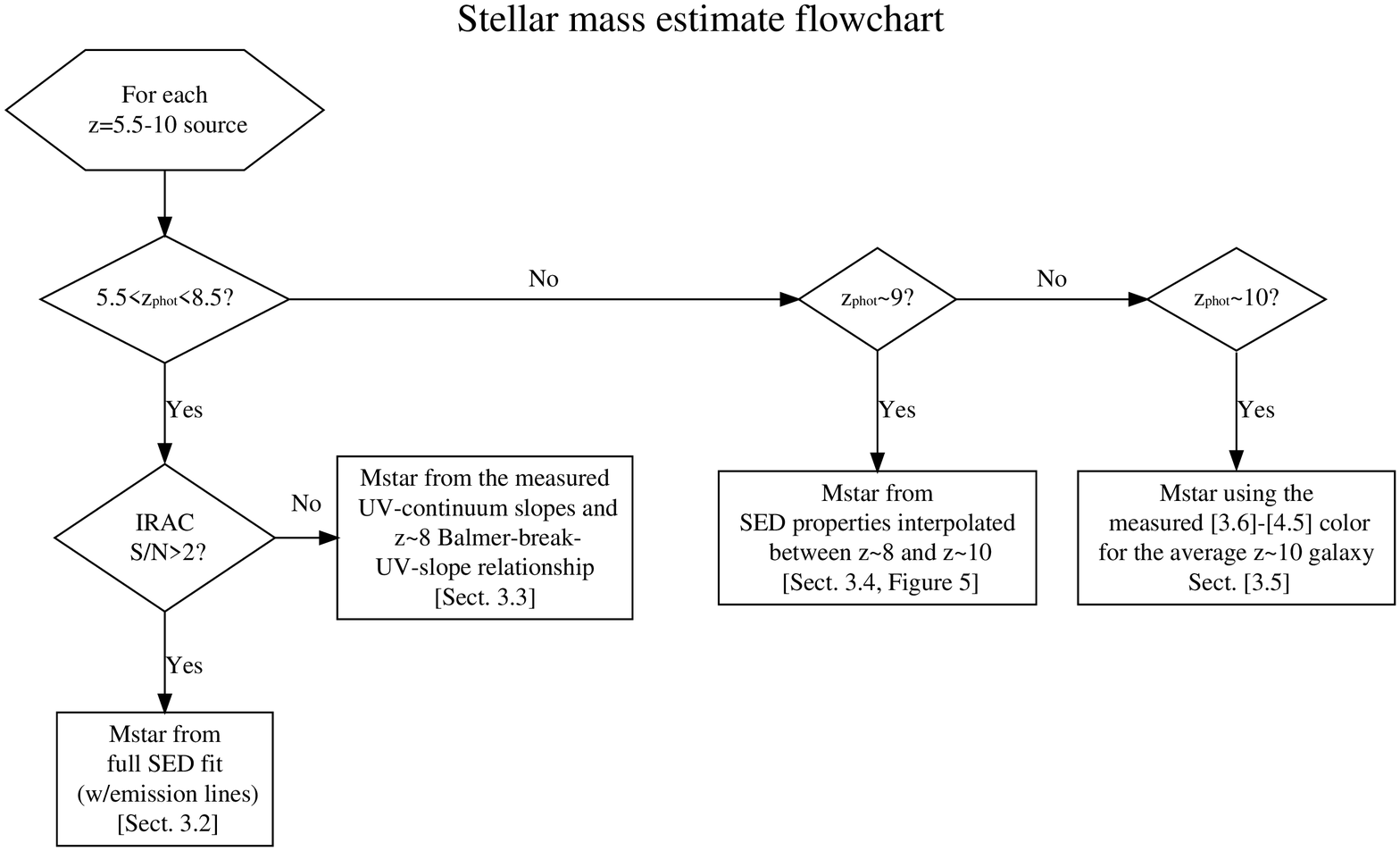}
\caption{Flow chart summarizing the different procedures followed to estimate the total stellar mass $\mathcal{M}_\star$ of galaxies in our sample depending on the source redshift and the significance of the IRAC detections. \label{fig:d_flux}}
\end{figure*}

\section{Stellar mass estimates}
\label{sect:mstar_0}
In this section we present the general framework adopted for estimating the stellar masses of the galaxies in our $z=6-10$ samples. The procedure we utilise for galaxies in our $z=6-8$ samples depends now whether we detect individual sources (at $2\sigma$ level or above) or not, to limit the impact of potential systematics. The two approaches are described in Sections \ref{sect:mstar_det} (IRAC-detected sources) and \ref{sect:mstar_undet} (IRAC non-detected sources). Furthermore, because the $3.6\mu$m band probes the rest-UV for $z\gtrsim9$, we implemented different procedures for the $z\sim9$ and $z\sim10$ samples, which we present in Section \ref{sect:mstar_z9} and \ref{sect:mstar_z10}, respectively. In Figure \ref{fig:d_flux} we present a flowchart to better understand the specific procedures adopted to compute the stellar mass of the sources in our samples, depending on redshift and significance of the IRAC detections for each individual source. 
  
\subsection{Modelling assumptions}
\label{sect:stellar_pop}
For our stellar population parameter estimates we considered the \citet{bruzual2003} composite stellar population models with a \citet{salpeter1955} initial mass function (IMF), a 0.2$Z_{\odot}$ metallicity and a constant star-formation history with a minimum age of $10^6$\,years and a maximum age set by the age of the Universe at each specific redshift. Template fitting was performed with  \textsc{FAST} \citep{kriek2009}, fixing the redshift of each source to the value produced by \textsc{EAzY}. In our fits, we consider a dust attenuation in the range $A_\mathrm{V}=0-3$\,mag with a \citet{calzetti2000} curve, assuming the same dust law for both the stellar continuum and the nebular emission.

Numerous studies suggest that the spectral energy distributions (SEDs) of galaxies observed at early epochs are characterized by strong nebular line emission (e.g., \citealt{schaerer2010, labbe2013,  stark2013, smit2014, debarros2019, faisst2016, faisst2019, faisst2020, endsley2021}), with typical equivalent widths EW$(H\alpha)$ and EW([\ion{O}{3}]+H$\beta$) in excess of few$\times100$\,\AA\ to $\sim1000$\,\AA. Furthermore, photoionization models predict that emission by nebular continuum could significantly contribute to the observed flux densities of young stellar populations even when they are probed through broad-band filters (e.g., \citealt{zackrisson2008, zackrisson2011, schaerer2010, inoue2011}).

We accounted for the contribution of nebular emission, both lines and continuum, processing the SED templates with \textsc{Cloudy} version 17.02 \citep{ferland2017}. For simplicity, we assumed a  spherical constant-density nebula with $n(H)=100$\,cm$^{-3}$, a gas metallicity matching that of the stellar component ($0.2 Z_\odot$), an ionization parameter $\log U=-2.5$, consistent with recent work (e.g., \citealt{stark2017,debarros2019}), and that the escape fraction was negligible.\\

We also implemented a second set of SED templates, where we added to the \citet{bruzual2003} templates only the effects of nebular continuum, ignoring any contribution from nebular line emission. This new set of SED templates was used in estimating stellar masses for those objects undetected in IRAC bands and sources at $z\sim9$ and $z\sim10$ after updating their IRAC flux densities using the phenomenologically-motivated relations described in Sections \ref{sect:mstar_undet}, \ref{sect:mstar_z9} and \ref{sect:mstar_z10}. An increasing number of studies indicate that the red IRAC colors observed for individual sources at $z\sim7-8$ could result from evolved stellar populations (e.g., \citealt{hashimoto2018, strait2020,roberts-borsani2020}). However, the observations unambiguously supporting such an interpretation regard just a few sources. The nebular line emission interpretation is supported by a recent study showing that on average $z\sim7-8$ LBGs have UV-optical colors consistent with no significant Balmer Break (\citealt{stefanon2021c}). Such Balmer-break sources therefore would not appear to have a large impact on the conclusions we draw regarding mass for statistical samples of $z\sim6-8$ galaxies. For these reasons, we only consider the color excess to be the result of contributions from nebular lines.

\subsection{Stellar mass estimates for sources detected by IRAC at $z\sim6$, $z\sim7$ and $z\sim8$}
\label{sect:mstar_det}

For the those sources in our $z\sim6, 7$, and $8$ samples with $\ge 2\sigma$ detections in both IRAC $3.6\mu$m and $4.5\mu$m bands, i.e., the majority of sources in these samples ($\sim 75\%$ - see e.g., Figure \ref{fig:sample}), we computed the stellar masses by running  \textsc{FAST} with the \citet{bruzual2003} template set enriched with nebular continuum and emission line information from \textsc{Cloudy}, obtained as described in Section \ref{sect:stellar_pop}, and shown in Figure \ref{fig:d_flux}.

\subsection{Stellar mass estimates for sources undetected by IRAC at $z\sim6$, $z\sim7$ and $z\sim8$}
\label{sect:mstar_undet}

Estimates of $\mathcal{M}_\star$ for the small fraction ($\sim25\%$) of galaxies in our sample that are not detected in IRAC are going to be quite uncertain by comparison. Fortunately,  we can make use of an observational correlation between the UV-continuum slope $\beta$ and the amplitude of the Balmer break that has been reported by both \citet{oesch2013} and \citet{stefanon2021c}. This correlation then provides a good proxy for the age of the stellar population.

Using both individual and stacked $z_\mathrm{phot} = 7.3 - 8.7$ sources, \citet{stefanon2021c} showed that there is a clear correlation between the UV-continuum slope $\beta$, that is determined using the $J$ and $H$ measurements, and the Balmer break amplitude.   For sources with the bluest UV slopes ($\beta\sim -2.5$), \citet{stefanon2021c} find blue ($H - [3.6] \sim -0.5$\, mag) colors, and the $H-[3.6]$ colors become increasing red as one moves to redder UV-continuum slopes $\beta\sim -1.6$ (see their Figure 5).  Because at these redshifts the $3.6\mu$m band probes rest-frame wavelengths just red-ward of the Balmer break, while the $H$ band probes the rest-frame UV, the above trend suggests that the UV slope could be used as a proxy for the break amplitude, and hence for the age of a stellar population.   \citet{oesch2013} found a very similar correlation between the amplitude of the Balmer break $H-[4.5]$ and the UV-continuum slope $\beta$ for $z\sim4$ galaxies.

To estimate new $3.6\mu$m flux densities, we therefore adopted the relationship between  $\beta$ and the $H-[3.6]$ color found by \citet{stefanon2021c} for $z\sim8$ LBGs, after correcting it for the effects ($0.2$ mag) of [\ion{O}{2}] emission contaminating the $3.6\mu$m band at $z\sim8$ (see \citealt{stefanon2021c} for more details):
\begin{align}
H-[3.6] &=0.03+1.78(\beta+2.2)   
\label{eq:h36}
\end{align}

This relationship was derived for galaxies with UV continuum slopes $\beta$ ranging from  $-2.6$  to $-1.9$. We adopted a constant value of $H-[3.6]=0.56$ mag when $\beta > -1.9$. The corresponding $4.5\mu$m flux densities were computed assuming the rest-frame optical had a flat $f_\nu$ SED. A flat $f_\nu$ SED is expected from stacking analysis of observations at similar redshifts (e.g., \citealt{gonzalez2012, stefanon2017b}) and it is predicted by photoionization modeling as the effect of nebular continuum emission in relatively young stellar populations of star-forming galaxies (e.g., \citealt{schaerer2009}). The flat SED hypothesis is also consistent with the negligible dust content found for $\mathcal{M}_\star\lesssim \mathcal{M}_\star^*$ galaxies at high redshifts (e.g., \citealt{bouwens2020}).

Our hypothesis of a flat SED at rest-frame optical wavelengths for $z\sim6-8$ is consistent with observations  only if we assume a negligible contribution of nebular lines in the $4.5\mu$m band.  Later in this section we describe how we accounted for this in our SED fitting. Following \citet{stefanon2021c}, we computed the UV slope $\beta$ from the best-fitting SED template of each individual source.

Having established the described correlation, the challenge became applying it to galaxies distributed over the redshift range $z\sim 6 - 8$, requiring that we account for the different rest-frame wavelengths of the $H_\mathrm{160}$ and $3.6\mu$m filters. To deal with this aspect, we updated the newly-computed $H-[3.6]$ color of each source assuming a flat $f_\nu$ SED at rest-frame optical wavelengths and a power-law with slope $\beta$ in the rest-UV. In doing so, we retained the same uncertainties for the IRAC flux measurements as originally estimated by \textsc{Mophongo}. The flux densities free of emission lines that we obtained from the above procedure were then used to derive our stellar mass measurements for those objects with S/N$<2$ in either one of the $3.6\mu$m or $4.5\mu$m bands. For this step, we ran \textsc{FAST} using the emission line-free template set. 

To test the robustness of the stellar mass measurements for the IRAC-undetected sources, we also computed the stellar mass of sources with $>2\sigma$ detection in both IRAC bands after replacing the IRAC flux densities with those obtained from Eq. \ref{eq:h36} and assuming $[4.5]=[3.6]$. In comparing stellar mass estimates made in these two separate ways, we recovered essentially identical results, validating this method.  These tests show that we can confidentially use this method for computing the stellar masses for $z\sim6, 7$ and $z\sim8$ galaxies detected at $<2\sigma$ significance with IRAC (see Appendix \ref{app:mstar_comp} for further details).

\subsection{Stellar mass estimates for the $z\sim9$ sample}
\label{sect:mstar_z9}

\begin{figure}
\hspace{-0.5cm}\includegraphics[width=9cm]{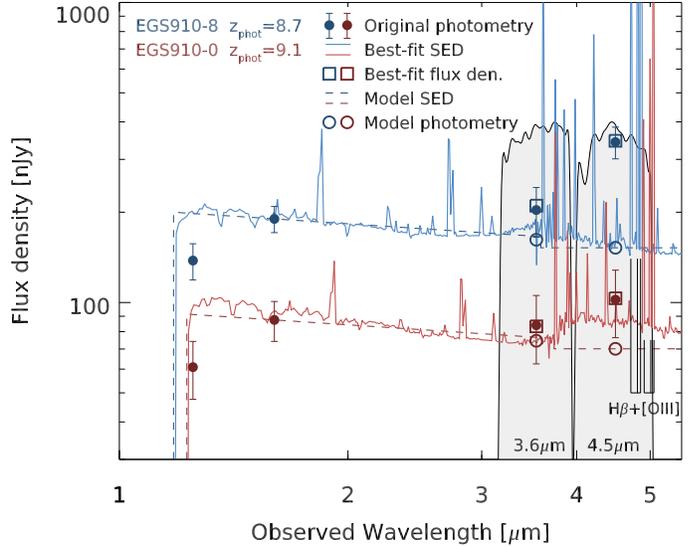}
\caption{Illustration of the challenges we face in estimating the stellar masses of $z \sim 9$ sources, given the uncertain position of the Balmer break/jump and [\ion{O}{3}]+H$\beta$ emission lines relative to the $3.6\mu$m and $4.5\mu$m bands, and of the procedure we adopted to reduce the associated systematic effects. Presented are two distinct sources in the upper and lower halves of the redshift range of our $z\sim9$ sample (EGS910-8 at $z_\mathrm{phot}\sim8.7$,  and EGS910-0  at $z_\mathrm{phot}\sim9.1$ - \citealt{bouwens2019a} - blue and red, respectively). All flux densities of EGS910-0 were arbitrarily rescaled by a factor $0.7$ to improve readability. The original photometric measurements are marked with filled circles, while the corresponding best-fit templates are presented as solid curves; the flux densities in the $3.6\mu$m and $4.5\mu$m bands for the best-fit templates are indicated as open squares. The two sets of vertical lines close to the bottom-right corner indicate the location of H$\beta$ and of the [\ion{O}{3}] doublet at the redshifts of the two galaxies. The two filled grey regions correspond to the transmission curves of the IRAC $3.6\mu$m and $4.5\mu$m bands, arbitrarily renormalized. The photometric redshift uncertainties ($\Delta z\sim0.6-0.8$) do not allow us to properly estimate the relative contribution of the rest-frame optical light to the observed flux density in the $3.6\mu$m band, and of some among the strongest emission lines (H$\beta$ and [\ion{O}{3}]) to the flux densities in the $4.5\mu$m band. These are necessary to constrain the amplitude of the Balmer Break and therefore the age of the stellar population. For each source, we therefore created a model SED (dashed curves), free from nebular lines contribution, interpolating the empirical relationships we derived from observations of $z\sim8$ and $z\sim10$ galaxies (see main text), and computed the expected flux densities in the IRAC $3.6\mu$m and $4.5\mu$m bands (open circles). We combined the new model photometry with the existing \textit{HST} measurements to estimate the stellar masses through a new SED fit (see Section \ref{sect:mstar_z9}). \label{fig:z9_fit_example}}
\end{figure}

The approaches as used in Section \ref{sect:mstar_det} and \ref{sect:mstar_undet} cannot be applied at $z\gtrsim9$. There are two limiting factors. First, complications arise from the Balmer break beginning to move into and through the IRAC $3.6\mu$m band at $z\sim9$, and second, it becomes more challenging to determine $\beta$ from the \textit{Hubble} WFC3 IR bands. More details of the challenges of working with $z\sim9$ galaxies are below, while in Figure \ref{fig:z9_fit_example} we present the application of our procedure to two of the sources in our $z\sim9$ sample.

First, the uncertainties in photometric redshifts (typical values of $\Delta z\sim0.6-0.8$) do not lead to clarity in the relative contribution from rest-frame UV (blueward of the Balmer break) and rest-frame optical light (redward of the Balmer break) to the IRAC $3.6\mu$m band (ranging from $\sim50\%$ optical light contribution at $z\sim8.7$ to $\lesssim10\%$ at $z\sim9.3$). Furthermore, the H$\beta$ and [\ion{O}{3}] emission lines can significantly contribute to the flux density in the $4.5\mu$m band up to $z\sim9.3$, potentially mimicking the existence of more evolved stellar populations. These make for an uncertain SED fit, limits insight into the break amplitude, and thereby increases the uncertainty in stellar ages and, consequently, stellar masses. Second, only the $JH_{140}$ and $H_{160}$ bands are available to probe the UV-continuum slopes for $z\sim9$ galaxies  (e.g., \citealt{dunlop2013, bouwens2014}). The wavelength coverage of this latter band substantially overlaps with that of $H_{160}$, limiting the wavelength leverage for the UV slope estimates, while its extension to the blue makes $JH_{140}$ sensitive to the intrusion of the Lyman Break at $z\sim9$, limiting its utility for measuring $\beta$. 

To overcome, at least in part, these challenges, we did not adopt the original IRAC photometry in our SED fitting, but instead we estimated the separate contributions above and below the Balmer break of the rest-frame UV and of the rest-frame optical light to the $3.6\mu$m band using the following procedure.

Supported by the consistent correlations found between the UV luminosity and the UV slopes $\beta$ for LBGs at $z>4$ (e.g., \citealt{bouwens2014b,  finkelstein2012, rogers2014, bhatawdekar2020}), we computed the  contribution to the $3.6\mu$m band from the rest-frame UV light assuming each source had a power-law-like SED with slope $\beta$ equal to the median of the UV slopes of $z\sim8$ galaxies in our sample with similar UV luminosity ($|\Delta M_\mathrm{UV}|\le0.5$\,mag).

The contribution from the rest-frame optical was then obtained by interpolating between the $z\sim8$ $H_\mathrm{160}-[3.6]$-$\beta$ relation (Equation \ref{eq:h36}) and the $H_\mathrm{160}-[4.5]$ color at $z\sim10$ (see Section \ref{sect:mstar_z10}) using the same median UV slopes adopted for the rest-UV light estimates.

These two contributions were ultimately combined, weighting by the corresponding fraction of the $3.6\mu$m-band coverage given the individual photometric redshifts. We also verified that $\sim94\%$ of the newly computed flux densities in the $3.6\mu$m band was consistent at $<2.4\sigma$ with with the original measurements, increasing the confidence on our procedure. The emission line-free flux density in the $4.5\mu$m band was then computed assuming the rest-frame optical has a flat $f_\nu$ SED. Stellar masses were finally obtained running \textsc{FAST}  on these reconstructed SEDs, adopting for consistency the SED template set where the emission lines have been explicitly removed. \\

Our $z\sim9$ sample includes one galaxy spectroscopically confirmed at $z_\mathrm{spec}=8.683$ (EGS910-10 - \citealt{roberts-borsani2016}). The spectroscopic redshift enables the unambiguous recovery of the contribution of the rest-frame optical light to the IRAC $3.6\mu$m band. An SED fit performed with the original photometry on the same template set adopted in Section \ref{sect:mstar_det} results in a stellar mass lower by only $0.05$\,dex than the value we obtained applying the procedure described above, increasing our confidence on the results. Finally, the median $H_\mathrm{160}-[3.6]$ color for the sources in our $z\sim9$ sample is $0.07$\,mag. This value is consistent with the $-0.03\pm0.14$ found at $z\sim8$ by \citet{stefanon2021c}, indicative of young stellar population ages, and supports the low $\mathcal{M}_\star/L_\mathrm{UV}$ values we find (see e.g., Table \ref{tab:mstar_muv}).

\subsection{Stellar mass estimates for the $z\sim10$ sample}
\label{sect:mstar_z10}

As we noted above, for $z>9$, the \emph{Spitzer}/IRAC $3.6\mu$m band begins to move blue-ward of the Balmer Break suggesting it could be effectively used for a UV-slope measurement.  However, despite the unprecedented depth provided by the GREATS mosaics, only $3/6$ of our $z\sim10$ candidates have $>2 \sigma$ detections in the $3.6\mu$m band. For these reasons before computing the stellar masses, and to guide the fitting, we updated the $3.6\mu$m and $4.5\mu$m flux densities of all sources with values corresponding to a flat SED, i.e., $H_\mathrm{160}-[3.6]=0.0$\,mag and blue $H_\mathrm{160}-[4.5]=-0.13$\,mag colors. These colors are consistent with the stacking analysis performed with the same sample of $z\sim10$ sources done by \citet{stefanon2021d}; moreover, an approximately flat UV slope at $z\sim10$ has been reported by  \citet{wilkins2016a} and, for $M_\mathrm{UV}\sim -21$\,mag sources at $z\sim9$, by \citet{bhatawdekar2020}. We reduced the $[4.5]$ flux density by $0.2$ mag to remove the estimated [\ion{O}{2}] contribution, resulting in an adopted $H_\mathrm{160}-[4.5]$ color of $-0.33$ mag for all sources. In the process, we maintained the same flux density uncertainties originally measured in the $3.6\mu$m and $4.5\mu$m bands. We verified that the $3.6\mu$m flux densities  of all sources computed in this way were consistent at $\sim2\sigma$ with the original measurements. This approach is consistent with computing the individual $\mathcal{M}_\star$ assuming all sources possess the same $\mathcal{M}_\star/L$ ratio, as derived from the stack of the $z\sim10$ sample.

\begin{figure*}
\includegraphics[width=18cm]{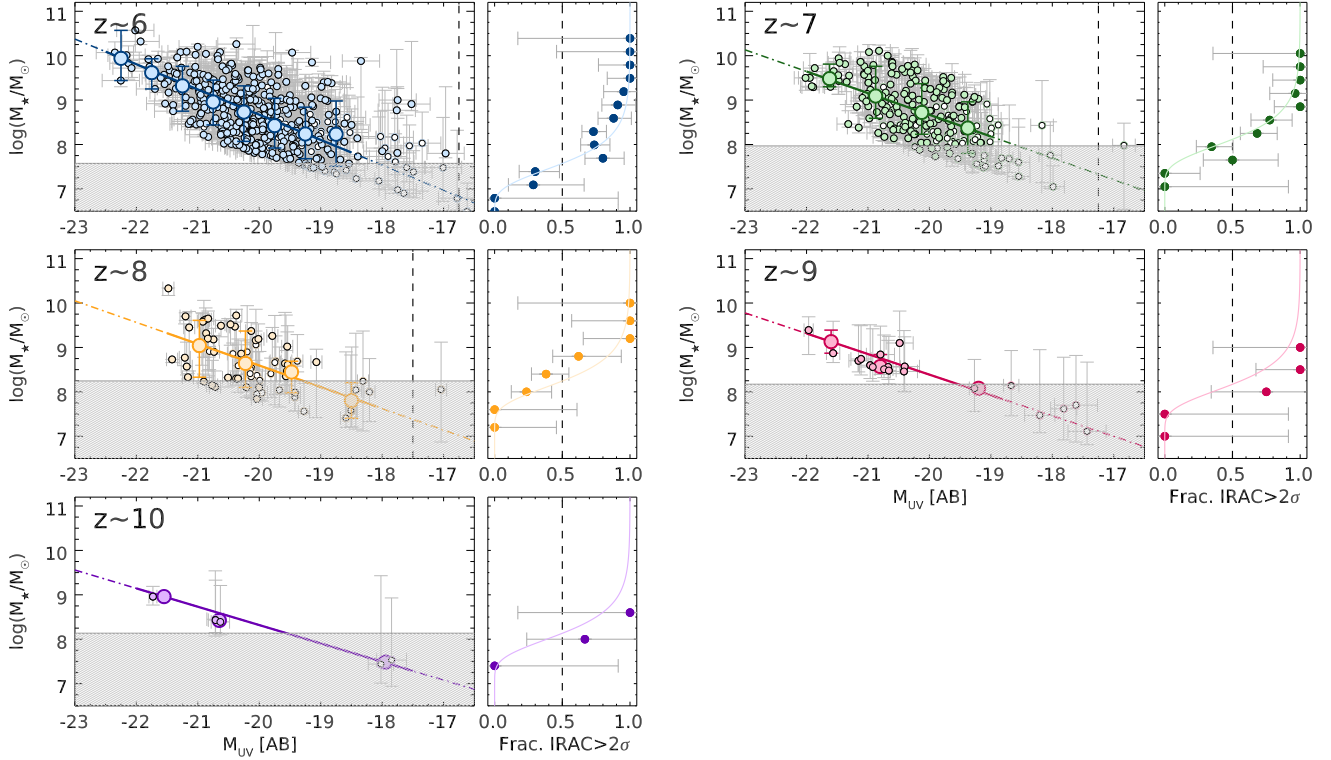}
\caption{For each redshift bin, indicated at the top-left corner of the larger panels, we present a set of two plots. In the righthand plot, we present the fraction of galaxies with $3.6\mu$m and $4.5\mu$m flux densities detected at $2\sigma$ level or better  ($4.5\mu$m only for the $z\sim10$ sample), in bins of stellar mass (colored points and error bars -- note the unusual plot orientation -- rotated by 90 degrees). The vertical dashed lines mark the $0.5$ fraction we adopted as our criterion to identify the lowest stellar mass that can confidently be used in the measurement of the stellar mass function,  while the colored solid curves mark the best-fitting Gompertz function (see text).  The plot on the left side of each panel shows the individual sources we selected in each redshift bin in the $M_\mathrm{UV}-\mathcal{M}_\star$ plane (open and filled small circles). Filled circles mark those sources detected at the $2\sigma$ level or better in both the $3.6\mu$m and $4.5\mu$m bands, while the small open circles correspond to sources with $S/N<2\sigma$ in at least one of the two IRAC bands. The large filled circles with errorbars correspond to the median and 68\% confidence intervals on the stellar masses in varying-width bins of UV luminosity. The colored solid lines indicate the best-fit linear relation through the median estimates, while the dot-dashed lines represents its extrapolation to brighter and fainter luminosities. For the $z\sim8, 9$ and $10$ redshift bins we also included median estimates for $M_\mathrm{UV}$ bins dominated by sources with  $<2\sigma$ in either IRAC band to better guide the fits. The vertical black dashed line marks the faintest UV luminosity down to which we can reliably recover the $z\sim6-8$ UV LF after removing sources with contaminated IRAC measurements (see Figure \ref{fig:LF_cc}); they are absent for the $z\sim9$ and $z\sim10$ bins because for these samples we did not apply any cleaning.  The hatched grey region identifies the range in stellar mass where the fraction of sources with $>2\sigma$ in both IRAC bands is smaller than  $0.5$. The sources selected for our SMF estimates have robust stellar mass estimates down to the applied limits; sample cleaning does not systematically affect the recovery of the volume densities down to $\mathcal{M}_\star\sim 10^8\mathcal{M}_\odot$ across the full range of redshifts considered here.  \label{fig:mstar-muv}}
\end{figure*}

\section{Completeness and selection biases}
\label{sect:completeness}

In Figure \ref{fig:mstar-muv} we present our sample in the stellar mass ($\mathcal{M}_\star$) vs. absolute UV magnitude ($M_\mathrm{UV}$) plane.  We indicate with open circles those sources with  $<2\sigma$ significance in either one of IRAC bands for the $z\sim6,7,8$ samples, or just in the $4.5\mu$m band for the $z\sim9$ and $z\sim10$ samples, providing a qualitative indication of the fraction of sources with more poorly-constrained stellar masses. The panels show an overall correlation between $\mathcal{M}_\star$ and $M_\mathrm{UV}$ at all redshifts, even though the scatter in $\mathcal{M}_\star$ can be as large as $\gtrsim1.5$\,dex  for specific $M_\mathrm{UV}$ values. Large scatter is predicted by some simulations, as the result of a real variation in the specific SFR (e.g., \citealt{ceverino2018}), but a detailed study of the mass-to-light ratios $\mathcal{M}_\star/L_\mathrm{UV}$ at these redshifts, while interesting, is beyond the scope of this work. 

For illustrative purposes in Figure \ref{fig:mstar-muv} we present a simple linear fit to the median values of the stellar masses in bins of $M_\mathrm{UV}$. In particular for this analysis, no statistical correction is made to account for our initial removal of sources with contamination by neighbours in the IRAC bands. The results of the fit are marked by colored lines in Figure \ref{fig:mstar-muv}, and are listed in Table \ref{tab:mstar_muv}. In fitting the linear relation at $z\sim8, 9$ and $10$, we included the measurements corresponding to $M_\mathrm{UV}\sim-18.5, -19.2$ and $-18$\,mag, respectively, dominated by sources undetected in IRAC (open circles in Figure \ref{fig:mstar-muv}), because the smaller sample size would otherwise make the fit very uncertain. This adds some arbitrariness to the slope estimates, but we judged that the linear fit would otherwise provide a worse representation of the individual measurements.   The slopes are consistent with a constant value of $\sim-0.55$ across the $z\sim6-8$ redshift range, and of $\sim-0.45$ at $z\sim9$ and $\sim10$.

In Appendix \ref{app:mstar-muv} and Figure \ref{fig:comp_mstar-muv} of Appendix \ref{app:mstar-muv} we include a more detailed comparison with previous work of our effective $\mathcal{M}_\star-M_\mathrm{UV}$ relation. Here we note that our slope estimates are in general consistent with previous determinations at similar redshifts (e.g., \citealt{duncan2014, song2016, bhatawdekar2019, kikuchihara2020}), while the intercept we have derived is lower on average by $\sim0.2-0.3$\,dex (see Figure \ref{fig:comp_mstar-muv} in Appendix \ref{app:mstar-muv}). Some such differences would not be unexpected, however, given that our deeper IRAC photometry provides more accurate insights into the stellar population properties of galaxies at $z\sim6-10$.\\

\begin{deluxetable}{cccc}
\tablecaption{$M_\mathrm{UV}$ vs. $\mathcal{M}_\star$  Linear Fit Parameters \label{tab:mstar_muv}}
\tablehead{\colhead{Redshift} &  \colhead{$\log\mathcal{M}_\star/\mathcal{M}_\odot$} & \colhead{Slope} & \colhead{$\mathcal{M}_\star/L_\mathrm{UV}$\tablenotemark{$\dagger$}}\\
\colhead{bin} & \colhead{for $M_\mathrm{UV}=-20.5$} & & \colhead{[$\mathcal{M_\odot}/L_\odot$]}
}
\startdata
$ 6$ & $ 9.0\pm 0.1$ & $-0.57\pm 0.02$ & $0.027^{+0.002}_{-0.002}$ \\
$ 7$ & $ 8.9\pm 0.1$ & $-0.49\pm 0.08$ & $0.024^{+0.004}_{-0.003}$ \\
$ 8$ & $ 8.8\pm 0.1$ & $-0.49\pm 0.12$ & $0.020^{+0.006}_{-0.005}$ \\
$ 9$ & $ 8.6\pm 0.1$ & $-0.46\pm 0.03$ & $0.012^{+0.001}_{-0.001}$ \\
$10$ & $ 8.5\pm 0.1$ & $-0.41\pm 0.02$ & $0.010^{+0.001}_{-0.001}$ \\
\enddata
\tablenotetext{\dagger}{Stellar mass-to-light ratio computed from the stellar mass for $M_\mathrm{UV}=-20.5$ mag. }
\end{deluxetable}

As expected, most of the sources at the faint, and typically low-mass, end only have marginal IRAC detections. The fraction of sources with IRAC detection significance in excess of $2\sigma$ is presented in the panels to the right of each  $\mathcal{M}_\star-M_\mathrm{UV}$ plot, where the error bars reflect the poissonian uncertainties. To represent analytically the dependence of IRAC detections with stellar mass, we fitted the following form of the Gompertz function:
\begin{equation}
f(\mathcal{M}_\star)=\exp[-m_0\exp(-a_0\mathcal{M}_\star)]
\end{equation}
where $a_0$ controls the steepness of the decrease in counts, while $m_0$ applies a rigid shift in stellar mass to the curve. The Gompertz function is a generalization of the logistic function, and it allows to approach the two asymptotes with different bendings (i.e., its shape is asymmetric). This constitutes a better representation of our measurements.  During the fit of the fractions for $z\ge7$, we only left $m_0$ free to vary, fixing $a_0$ to the value obtained at $z\sim6$ ($a_0\equiv2.4$). Note that fitting for both parameters at $z\sim7$ produced a value of $a_0$ very similar to that of $z\sim6$, although with larger uncertainties.

We considered our SMF to be robustly determined for stellar masses larger than those corresponding to a fraction $f(\mathcal{M}_\star)=0.5$, that we consider as a fair ratio between sources with robust stellar mass determinations and those with more unconstrained estimates. However, given the relatively steep slope found for the Gompertz functions, even the bins of the SMF corresponding to  the lowest stellar masses will contain a fraction of sources with detections in both IRAC bands $>0.5$.  With this $0.5$ limit, inverting the Gompertz functions result in lower bounds of $\log(\mathcal{M}_\star/\mathcal{M}_\odot)=7.6, 8.0, 8.2, 8.2$ and $8.1$ at $z\sim6, 7, 8, 9$ and $z\sim10$, respectively.

Because of the correlation between the UV luminosity and stellar mass, one may expect that our ability to reconstruct the UV LF only to absolute magnitudes $\sim1-2$\,mag brighter than the detection limits (see our discussion in Section \ref{sect:sample} and  Appendix \ref{app:LF_cc}) could systematically affect our measurements of the stellar mass function by excluding otherwise legitimate sources. However, Figure \ref{fig:mstar-muv} shows that the limits in UV luminosity that we find do not impact the stellar mass completeness of our $z\sim6$ samples, and only very marginally impact those at $z\sim7$ and $z\sim8$, where the selection in UV luminosity excludes just two sources close to our stellar mass threshold. Since their stellar mass is constrained to only within $\sim1$\,dex, their value for our sample is minimal. Overall, these results increase our confidence in the sample selection and stellar mass function measurements.\\

Our sample selection relies on Lyman-break criteria, which, by construction, are biased against evolved, redder systems, more likely included in selections exclusively based on photometric redshift criteria (see also \citealt{fontana2006, duncan2014, grazian2015, song2016, stefanon2017b}). In \citet{stefanon2017b} we showed that at $z\sim4$ the LBG criteria are able to recover at least $\sim75\%$ of the sources from photometric redshift selections for stellar masses $\mathcal{M}_\star\lesssim10^{10}M_\odot$ and concluded that the $z\sim4$ sample was only marginally affected by LBG selection criteria compared to photometric redshift ones. Because we expect that the fraction of evolved systems at $z>4$ is even lower than at $z\sim4$, we consider the effects of LBG selection marginal compared to photometric redshifts selections. 

Interestingly, an increasing number of studies (e.g., \citealt{yan2004, huang2011, caputi2012, caputi2015, stefanon2015, wang2016, wang2019, williams2019, alcalde-pampliega2019, fudamoto2020, gruppioni2020}) are revealing the existence of extremely red, massive objects with $\mathcal{M}_\star\gtrsim10^{10.5}\mathcal{M}_\odot$ at $z>3-4$ (see, e.g., \citealt{wang2019, alcalde-pampliega2019}).  Most of these would remain hidden at higher redshifts, even at NIR wavelengths usually adopted for the detection of high-redshift sources in deep extragalactic fields. However, the limited samples and poor knowledge of their physical properties make estimating their contribution to the stellar mass budget at higher redshifts highly uncertain.  Nonetheless, we expect, as noted above, that their integral contribution to the SMF will overall be small, even if they may contribute more at the highest masses.

While we have mentioned different approaches to sample selection, we note that the Lyman-break criteria we use constitute a set of well-defined color selections that can be modelled and univocally reproduced when accurate comparisons are needed. The Lyman-break approach thus largely reduces any impact of selection biases that would affect our derivation of the characteristics of the intrinsic population of galaxies.

\section{Results}
\label{sect:results}
\subsection{The Stellar Mass Functions at $z\sim6-10$}
\label{sect:SMF}

We measured the SMF in bins of redshift centered at $z\sim6, 7, 8, 9$ and $z\sim10$, using the $V_\mathrm{max}$ estimator of \citet{avni1980}, which allows us to self-consistently combine samples selected from data of different depths. We adopted the co-moving volumes of \citet{bouwens2015}, which already account for the effects of detection incompleteness, LBG selection and photometric redshift scatter. Uncertainties were computed with the binomial approximation of \citet{gehrels1986}, adding in quadrature  cosmic variance from \citet{moster2011}, consistent with more recent determinations (e.g., \citealt{bhowmick2020}), after rescaling it by the square root of the number of fields (e.g., \citealt{driver2010}).

\begin{figure}
\includegraphics[width=9cm]{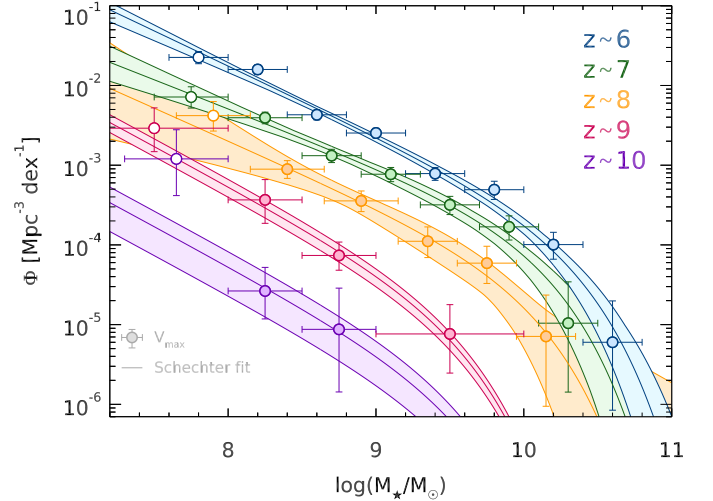}
\caption{The colored circles with error bars correspond to our $V_\mathrm{max}$ estimates of the SMF at $z\sim6$ to $\sim10$, following the color scheme presented in the legend at the top-right corner of the figure. The open circles at the lowest masses identify those measurements corresponding to stellar masses below our confidence threshold (Section \ref{sect:completeness}). These low-mass points have large, uncertain corrections. The solid colored curves mark the best-fitting Schechter functions, while the filled areas show those regions preferred at $68\%$ confidence. \label{fig:SMF}}
\end{figure}

The resulting measurements are listed in Table \ref{tab:SMF} and presented in Figure \ref{fig:SMF}.    Remarkably, the volume density of galaxies with stellar mass $\log(\mathcal{M}_\star/\mathcal{M}_\odot)\sim8.8$ increased by about $3$ orders-of-magnitude in the $\sim500$\,Myr elapsed between $z\sim10$ and $z\sim6$, suggesting an extremely rapid growth of the total stellar mass in galaxies at such early epochs.

\begin{deluxetable}{ccc}
\tablecaption{V$_\mathrm{max}$ determinations of the SMF \label{tab:SMF}}
\tablehead{\colhead{Redshift bin} & \colhead{$\log(\mathcal{M}_\star/\mathcal{M}_\odot)$\tablenotemark{a}} & \colhead{$\phi$}  \\
\colhead{} & \colhead{} & \colhead{[$ \times 10^{-4} \mathrm{dex}^{-1} \mathrm{Mpc}^{-3} $]} 
}
\startdata
$ 6$ & $ 7.80\pm0.20$\tablenotemark{$\dagger$} & $225^{+42}_{-37}$\tablenotemark{$\dagger$} \\
     & $ 8.20\pm0.20$ & $159^{+23}_{-21}$ \\
     & $ 8.60\pm0.20$ & $42.9^{+6.3}_{-5.9}$ \\
     & $ 9.00\pm0.20$ & $25.3^{+3.8}_{-3.5}$ \\
     & $ 9.40\pm0.20$ & $ 7.85^{+1.50}_{-1.40}$ \\
     & $ 9.80\pm0.20$ & $ 4.93^{+1.36}_{-1.21}$ \\
     & $10.20\pm0.20$ & $ 1.01^{+0.43}_{-0.35}$ \\
     & $10.60\pm0.20$ & $ 0.0601^{+0.1381}_{-0.0517}$ \\
 & & \\
$ 7$ & $ 7.75\pm0.25$\tablenotemark{$\dagger$} & $71.7^{+23.7}_{-18.3}$\tablenotemark{$\dagger$} \\
     & $ 8.25\pm0.25$ & $39.4^{+6.9}_{-6.3}$ \\
     & $ 8.70\pm0.20$ & $13.2^{+2.7}_{-2.4}$ \\
     & $ 9.10\pm0.20$ & $ 7.70^{+1.67}_{-1.49}$ \\
     & $ 9.50\pm0.20$ & $ 3.18^{+0.88}_{-0.78}$ \\
     & $ 9.90\pm0.20$ & $ 1.68^{+0.63}_{-0.53}$ \\
     & $10.30\pm0.20$ & $ 0.104^{+0.240}_{-0.090}$ \\
 & & \\
$ 8$ & $ 7.90\pm0.25$\tablenotemark{$\dagger$} & $ 41.9^{+20.6}_{-14.5}$\tablenotemark{$\dagger$} \\
     & $ 8.40\pm0.25$ & $ 8.91^{+2.49}_{-2.08}$ \\
     & $ 8.90\pm0.25$ & $ 3.56^{+1.19}_{-0.95}$ \\
     & $ 9.35\pm0.20$ & $ 1.11^{+0.57}_{-0.42}$ \\
     & $ 9.75\pm0.20$ & $ 0.591^{+0.371}_{-0.262}$ \\
     & $10.15\pm0.20$ & $ 0.0711^{+0.1637}_{-0.0617}$ \\
 & & \\
$ 9$\tablenotemark{$\ddagger$} & $ 7.50\pm0.50$\tablenotemark{$\dagger$} & $29.1^{+23.0}_{-13.9}$\tablenotemark{$\dagger$} \\
     & $ 8.25\pm0.25$ & $ 3.67^{+2.93}_{-1.81}$ \\
     & $ 8.75\pm0.25$ & $ 0.738^{+0.348}_{-0.256}$ \\
     & $ 9.50\pm0.50$ & $ 0.0764^{+0.1016}_{-0.0517}$ \\
 & & \\
$10$ & $ 7.65\pm0.35$\tablenotemark{$\dagger$} & $12.0^{+15.8}_{-7.8}$\tablenotemark{$\dagger$} \\
     & $ 8.25\pm0.25$ & $ 0.264^{+0.258}_{-0.146}$ \\
     & $ 8.75\pm0.25$ & $ 0.0872^{+0.1997}_{-0.0729}$ \\
 & & \\
\enddata
\tablenotemark{a}{Central value and range of each stellar mass bin}
\tablenotetext{\dagger}{This mass bin is dominated by sources with S/N$<2$ in either IRAC bands and lies below our fiducial completeness threshold (Section \ref{sect:completeness}), making the corresponding volume density very uncertain (open points in Figure~\ref{fig:SMF}).}
\tablenotetext{\ddagger}{Stellar mass estimates at $z\sim9$ are particularly challenging to constrain with current observations because the uncertainties in photometric redshifts do not allow us to ascertain where the $3.6\mu$m band lies relative to the Balmer Break, i.e., whether contributions to the $3.6\mu$m band are primarily the rest-frame UV or the rest-frame optical or a combination of the two. A separate but similar challenge for $z\sim9$ galaxies is the lack of knowledge as to the degree to which the $4.5\mu$m band is contaminated by strong nebular line emission.}
\end{deluxetable}

\begin{figure*}
\includegraphics[width=18cm]{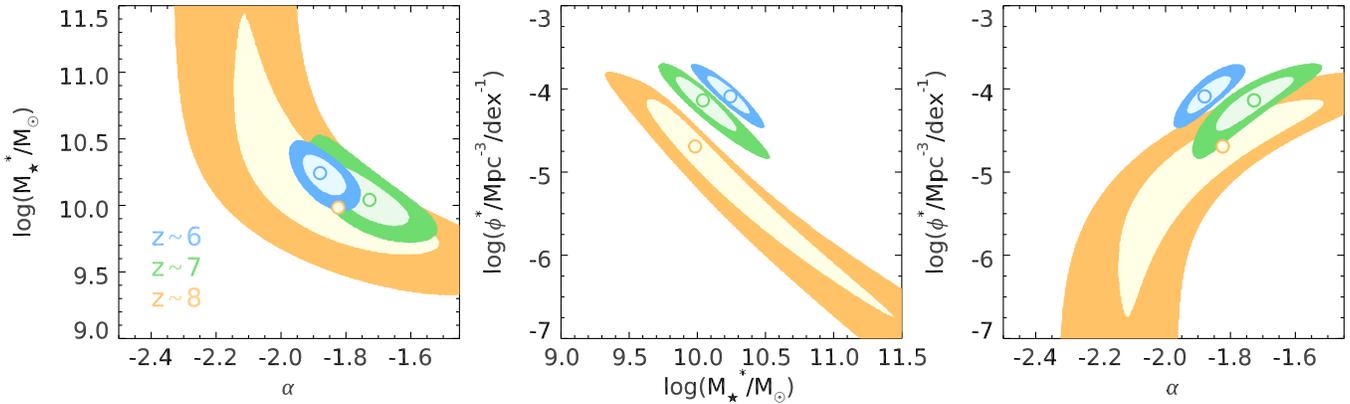}
\caption{68\% and 95\%  confidence intervals (light- and dark-shaded contours, respectively) for the best-fit Schechter parameters describing our SMFs at $z\sim6$, $z\sim7$ and $z\sim8$ (blue, green and orange contours, respectively). The colored circles mark the best-fit values at each redshift. The large uncertainties associated with the Schechter parameters prevent any unambiguous assessment of the evolution of the SMF from $z\sim8$ to $z\sim6$; however, there is clear evidence for a positive evolution in the characteristic stellar mass $\mathcal{M}^*_\star$ and normalization factor $\phi^*$ with cosmic time. \label{fig:chi2}}
\end{figure*}

\begin{figure}
\includegraphics[width=8cm]{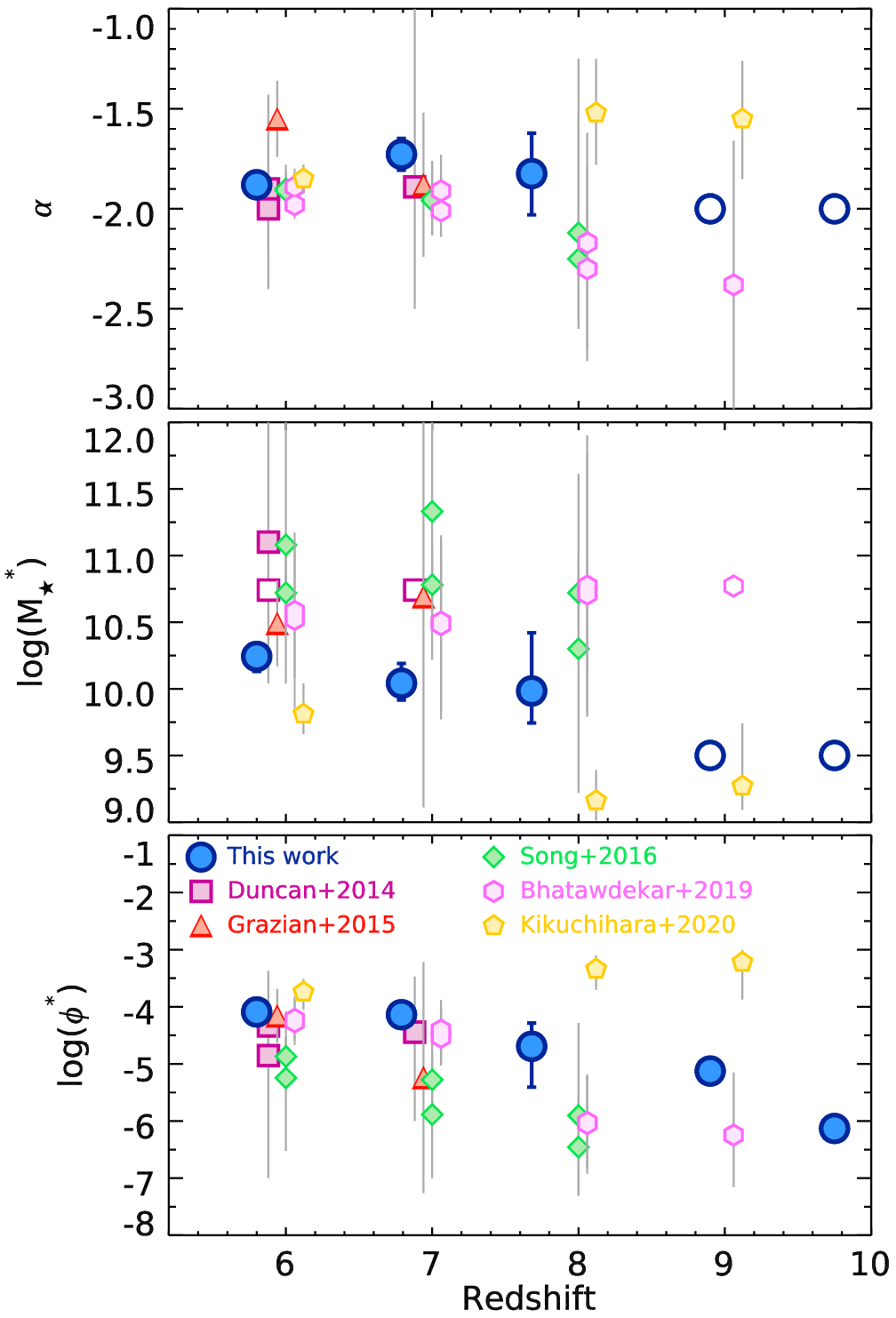}
\caption{Evolution we derive for $\phi^*$,$\mathcal{M}^*_\star$, and $\alpha$ from  $z\sim10$ to $z\sim6$ and a comparison to other determinations of these parameters from \citet[magenta squares]{duncan2014}, \citet[red triangles]{grazian2015}, \citet[green diamond]{song2016}, \citet[pink hexagons]{bhatawdekar2019}, and \citet[yellow pentagons]{kikuchihara2020} over the redshift range $z\sim6-10$. Open symbols show cases where some parameter values were kept fixed during the fit (see text).  \label{fig:comp_schechter}}
\end{figure}

We used the $V_\mathrm{max}$ measurements to fit a \citet{schechter1976} functional form, whose expression for logarithmic stellar masses  is:
\begin{equation}
\phi(m)dm=\ln(10)\phi^*10^{(m-m^*)(1+\alpha)}\exp(-10^{(m-m^*)})dm
\end{equation}

\noindent where $m=\log(\mathcal{M}_\star/\mathcal{M}_\odot)$, $m^*=\log(\mathcal{M}^*_\star/\mathcal{M}_\odot)$, with $\mathcal{M}^*_\star$ being the pivot mass between the power law and the exponential regimes, $\alpha$ corresponds to the slope at the low-mass end, and $\phi^*$ is the overall normalization factor. For the $z\sim6, 7$ and $z\sim8$ redshift bins we allowed all three parameters to vary, while for the $z\sim9$ and $z\sim10$ bins we kept the low-mass end slope $\alpha$ and characteristic mass $\mathcal{M_\star^*}$ fixed to $\alpha\equiv -2$ and $\log(\mathcal{M}_\star^*/\mathcal{M}_\odot)=9.5$\footnote{If we let $\mathcal{M}_\star^*$ vary in the $z\sim9$ fit, we obtain $\mathcal{M}_\star^*=10^{9.54}M_\odot$.}, respectively. The resulting parameterizations are represented by solid curves in Figure \ref{fig:SMF}, with the filled areas showing the $68\%$ confidence regions on the three parameters obtained by Monte Carlo sampling the Schechter parameterizations.  The values of the Schechter parameters and their $68\%$ uncertainties are listed in Table \ref{tab:schechter}, while Figure \ref{fig:chi2} presents the contours corresponding to the $68\%$ and $95\%$ confidence regions where $\Delta\chi^2\le2.30$ and $6.18$, respectively.

Unsurprisingly, the confidence intervals in Figure \ref{fig:chi2} show that there is a considerable range in $\phi^*$,$\mathcal{M}^*_\star$, and $\alpha$ values that reasonably represent the observed mass functions -- which is a reflection of how covariant the Schechter parameters are. Fortunately, in the case of the evolution of the characteristic stellar mass and the number density normalization factor (featured in the central panel of Figure \ref{fig:chi2}), the evolution is much clearer thanks to its being mostly orthogonal to the degeneracy between the two parameters. A similar result was found by e.g. \citet{grazian2015} for $z\sim4-7$. This suggests that the evolution of the SMF between $z\sim8$ and $z\sim6$ proceeded both in stellar mass and in number density.

Our measurements suggest a constant slope $\alpha\sim-1.8$ between $z\sim6$ and $z\sim8$, and are generally consistent at $\lesssim2\sigma$ with previous results (Figure \ref{fig:comp_schechter} - e.g., \citealt{duncan2014, song2016, bhatawdekar2019} and  \citealt{kikuchihara2020}). Our characteristic stellar masses are lower by $\sim0.2-0.5$\,dex compared to previous results at  $z\sim6-8$.  This may be due to the lower number density of the most massive galaxies that we find compared to the literature, supported by the unique deep wide-area coverage at rest-frame optical. However, robustly constraining the massive end of the SMF would require combining datasets with depth and area similar to what considered in this study with others covering $\gtrsim10\times$ larger areas than are currently available. Significant progress towards this end has been recently made by the SMUVS (\citealt{caputi2017,ashby2018}), COMPLETE (PI: Labb\'e), and COMPLETE2 (PI: Stefanon) programs, covering with deep ($\gtrsim 40$\,hr, corresponding to $1\sigma$ nominal sensitivity of $\sim30$\,nJy) \textit{Spitzer}/IRAC data a large parch of sky ($\sim1$degree$^2$) centered on the COSMOS/UltraVISTA footprint.  Finally, our estimate of the number density normalization parameter suggests a smooth decrease with increasing redshift, similar to what is seen by others. 

The Schechter parameterizations are useful for a synthetic representation of the SMFs. In the next section, we compare our $V_\mathrm{max}$ measurements to the corresponding ones from the literature.

\begin{deluxetable}{cccc}
\tablecaption{SMF Schechter fit parameters \label{tab:schechter}}
\tablehead{\colhead{Redshift bin} & \colhead{$\alpha$} & \colhead{$\log(\mathcal{M}^*_\star/\mathcal{M}_\odot)$} & \colhead{$\log(\phi^*/ \mathrm{dex}^{-1}/ \mathrm{Mpc}^{-3})$}  
}
\startdata
$ 6$ & $-1.88^{+0.06}_{-0.03}$ & $10.24^{+0.08}_{-0.11}$ & $-4.09^{+0.17}_{-0.12}$ \\
$ 7$ & $-1.73^{+0.08}_{-0.08}$ & $10.04^{+0.15}_{-0.13}$ & $-4.14^{+0.19}_{-0.23}$ \\
$ 8$ & $-1.82^{+0.20}_{-0.21}$ & $ 9.98^{+0.44}_{-0.24}$ & $-4.69^{+0.40}_{-0.72}$ \\
$ 9$ & $-2.00$ [fixed] & $ 9.50$ [fixed] & $-5.12^{+0.10}_{-0.13}$ \\
$10$ & $-2.00$ [fixed] & $ 9.50$ [fixed] & $-6.13^{+0.19}_{-0.36}$ \\
\enddata
\tablenotemark{}{Letting $\mathcal{M}_\star^*$ vary in the $z\sim9$ fit gives a very similar $\mathcal{M}_\star^*=10^{9.54}M_\odot$.}
\end{deluxetable}

\begin{figure*}
\includegraphics[width=18cm]{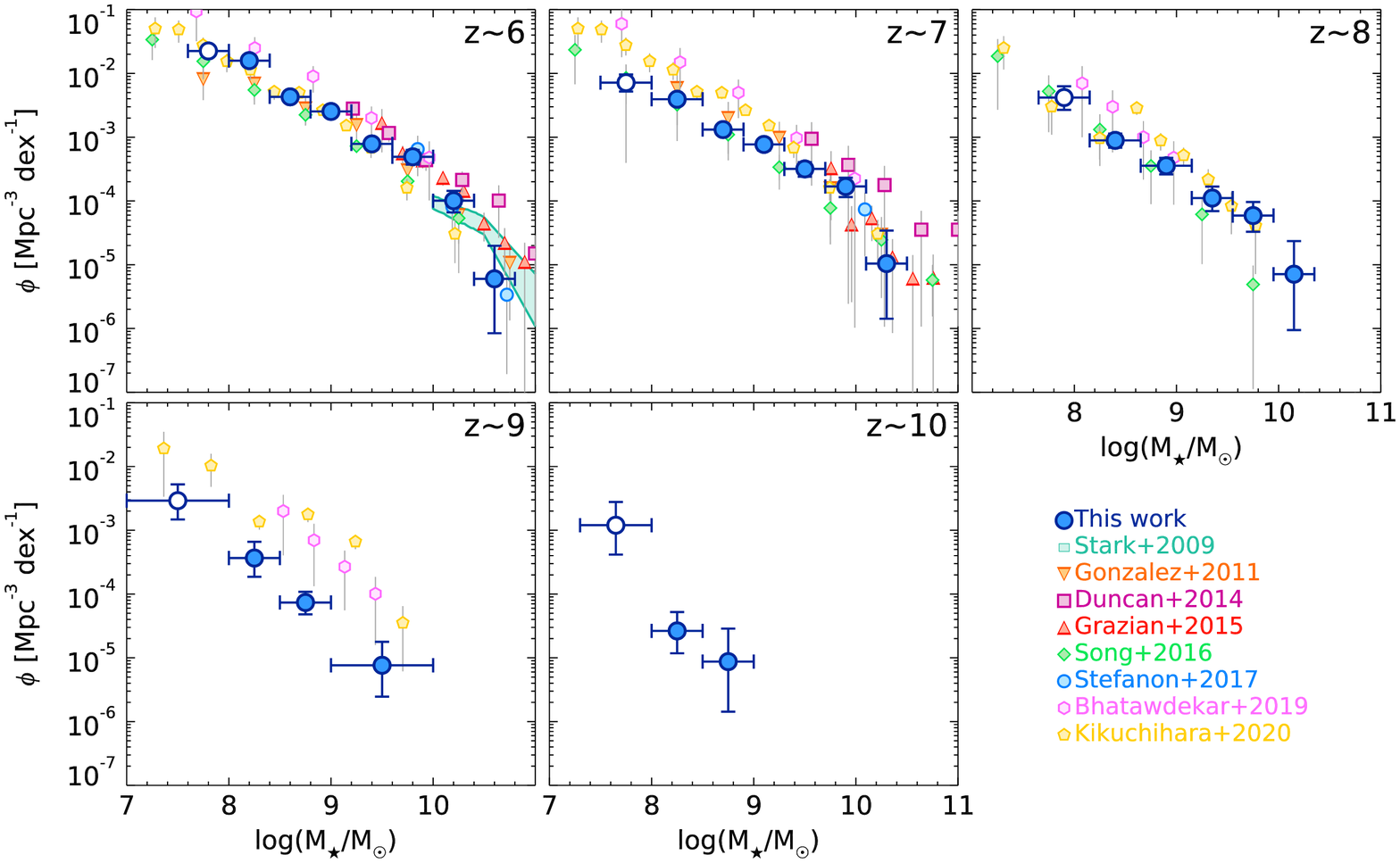}
\caption{Comparison of the SMF estimates from this work to previous determinations. Specifically, we here consider the SMFs of \citet{stark2009, gonzalez2011, grazian2015, duncan2015, song2016, stefanon2017b,bhatawdekar2019} and \citet{kikuchihara2020}, as listed in the legend in the lower right. We converted those stellar masses based on the \citet{chabrier2003} IMF into a \citet{salpeter1955} IMF through a 1.7 multiplicative factor. The redshift for each SMF panel is indicated in the top-right corner. \label{fig:smf_comparison}}
\end{figure*}

\subsection{Comparison to previous estimates}
\label{sect:comparison}

In Figure \ref{fig:smf_comparison} we compare our $V_\mathrm{max}$ estimates to previous studies. The large number of sources in our SMFs from the larger HST dataset, combined with the much deeper IRAC GREATS dataset, enabled us to generate SMFs at $z\sim6-10$ that have robust rest-frame optical underpinnings and improved sample statistics. Specifically we show the results of \citet{stark2009, gonzalez2011, duncan2014, grazian2015, song2016, stefanon2017b, bhatawdekar2019} and \citet{kikuchihara2020}.  We applied a factor of 1.7 (\citealt{madau2014}) to convert those stellar masses originally computed with a \citet{chabrier2003} IMF into a \citet{salpeter1955} IMF.

At $z\sim6$, for $\mathcal{M}_\star>5\times10^9M_\odot$ our SMF is consistent at $\sim1\sigma$ with the estimates of \citet{stark2009, gonzalez2011, grazian2015, stefanon2017b, bhatawdekar2019} and \citet{kikuchihara2020}, while there is a somewhat larger $\gtrsim2\sigma$ tension with the measurements of \citet{duncan2014}. Our low-mass end is consistent with the SMF of \citet{gonzalez2011, song2016} and \citet{kikuchihara2020}, with marginal indication of lower volume densities than \citet{bhatawdekar2019}, and more consistent with \citet{duncan2014}. At $z\sim7$ and $z\sim8$, there is an increased scatter among the measurements in the literature and larger uncertainties. Our $z\sim7$ SMF estimate lies amongst the current measurements for $\mathcal{M}_\star\gtrsim10^9M_\odot$, while for lower stellar masses,  our SMF lies closer to the measurements of \citet{gonzalez2011} and \citet{song2016}. At $z\sim8$ our measurements at  $\mathcal{M}_\star\lesssim10^9M_\odot$ also fall among the estimates of  \citet{song2016, bhatawdekar2019} and \citet{kikuchihara2020}, while at higher stellar masses the measurements of \citet{kikuchihara2020} are more consistent with our results. 

At $z\sim9$ our $\mathcal{M}_\star\sim10^9M_\odot$ measurement is lower by $\sim1$\,dex than other recent estimates of \citet{bhatawdekar2019} and \citet{kikuchihara2020}, while our measurement at higher mass is consistent with that of \citet{kikuchihara2020}. We remind the reader that estimating stellar masses at $z\sim9$ is very difficult because the large uncertainties in photometric redshifts do not properly allow us to ascertain where the $3.6\mu$m band falls relative to the Balmer Break, and so we cannot define the relative contributions above ("optical") and below ("UV") the Break. At $z\sim9$ in the $4.5\mu$m band there is the additional challenge of unknown levels of contamination by the strong nebular lines ([\ion{O}{2}] or [\ion{O}{3}]+H$\beta$). The discrepancies observed at $z\sim 9$ should therefore be viewed with more caution.  We are not aware of any other SMF estimates based on samples at $z\sim10$.

\section{Discussion}
\label{sect:discussion}

\subsection{Dispersion of SMF measurements}

The compilation of estimates presented in Figure \ref{fig:smf_comparison} shows an overall good agreement. However, for some redshifts and/or mass bins systematic differences can be as large as $\sim0.5-1.0$\,dex. A thorough analysis of the causes of these systematic differences goes beyond the scope of this paper. Here we limit our comments to briefly outline some possible causes. 

A first potential effect could be systematics in stellar mass estimates from different studies as a result of the different assumptions on the star-formation history (e.g., constant versus exponential or delayed SFH, e.g., \citealt{michalowski2014, mobasher2015,  leja2019, lower2020}) and nebular emission (e.g., \citealt{stark2013}). Furthermore, as already discussed by \citet{duncan2014}, photometric redshift selected samples (e.g., \citealt{duncan2014, grazian2015, song2016,bhatawdekar2019}) potentially include redder sources which are usually excluded by the LBG criteria (e.g., \citealt{stark2009,gonzalez2011,stefanon2017b,kikuchihara2020}), resulting in higher volume densities for the stellar mass functions of photometric redshift-selected samples. However, this is not always the case here: for instance, the photometric redshift-selected SMF at $z\sim6$ of \citet{song2016} is consistently lower than the LBG-based SMF of \citet{kikuchihara2020}. One further possibility was discussed by \citet{vulcani2017}, and results from contamination of LBG samples by lower-redshift interlopers. This particularly affects faint sources lacking sufficiently deep imaging at wavelengths bluer than the nominal Lyman break. Finally, cosmic variance may have significant impact  at the highest redshifts (e.g., \citealt{bhowmick2020}) because the photometric depth necessary to constrain the samples are only available over small ($\lesssim 100$arcmin$^2$) areas.

\subsection{Stellar mass density}
\label{sect:SMD}

\begin{deluxetable*}{cccccc}
\tablecaption{Stellar and dark matter halo mass densities \label{tab:SMD}}
\tablehead{\colhead{Nominal redshift} & \colhead{Median redshift} & \colhead{$\log(\rho_\star/\mathcal{M}_\odot/\mathrm{Mpc}^{-3})$\tablenotemark{a}} & \colhead{$\log(\rho_\mathrm{h}/\mathcal{M}_\odot/\mathrm{Mpc}^{-3})$\tablenotemark{b}} & \colhead{$\mathcal{\log(M}_\mathrm{h,lim}/\mathcal{M}_\odot)$\tablenotemark{c}} & \colhead{$\log(\rho_\mathrm{h}/\rho_\star)$\tablenotemark{d}} 
}
\startdata
$ 6$ & $ 5.80$ & $  6.68^{ +0.09}_{ -0.11}$ & $  8.64^{ +0.10}_{ -0.03}$ & $10.46^{+0.05}_{-0.11}$ & $  1.95^{ +0.15}_{ -0.10}$ \\
$ 7$ & $ 6.79$ & $  6.26^{ +0.13}_{ -0.17}$ & $  8.20^{ +0.09}_{ -0.05}$ & $10.54^{+0.05}_{-0.09}$ & $  1.93^{ +0.19}_{ -0.14}$ \\
$ 8$ & $ 7.68$ & $  5.73^{ +0.21}_{ -0.33}$ & $  7.69^{ +0.11}_{ -0.06}$ & $10.55^{+0.08}_{-0.10}$ & $  1.96^{ +0.35}_{ -0.22}$ \\
$ 9$ & $ 8.90$ & $  4.89^{ +0.25}_{ -0.29}$ & $  6.97^{ +0.20}_{ -0.16}$ & $10.47^{+0.09}_{-0.12}$ & $  2.08^{ +0.36}_{ -0.30}$ \\
$10$ & $ 9.75$ & $  3.68^{ +0.52}_{ -0.79}$ & $  6.04^{ +0.20}_{ -0.22}$ & $10.72^{+0.08}_{-0.10}$ & $  2.35^{ +0.81}_{ -0.56}$ \\
\enddata
\tablenotetext{a}{Logarithm of the stellar mass density, obtained integrating the SMF down to a stellar mass limit of $\mathcal{M}_\star=10^8\mathcal{M}_\odot$.}
\tablenotetext{b}{Logarithm of the dark matter halo mass density, obtained by integrating the HMF down to a halo mass limit of $\mathcal{M}_\mathrm{h}=\mathcal{M}_\mathrm{h,lim}$ (see note $c$ below).}
\tablenotetext{c}{Logarithm of the halo mass obtained from our abundance matching procedure, used as a lower limit in the measurement of the halo mass density}
\tablenotetext{d}{Ratio between the measured halo and the stellar mass densities, in log units.}
\end{deluxetable*}

\begin{figure*}
\hspace{-0.6cm}\includegraphics[width=18cm]{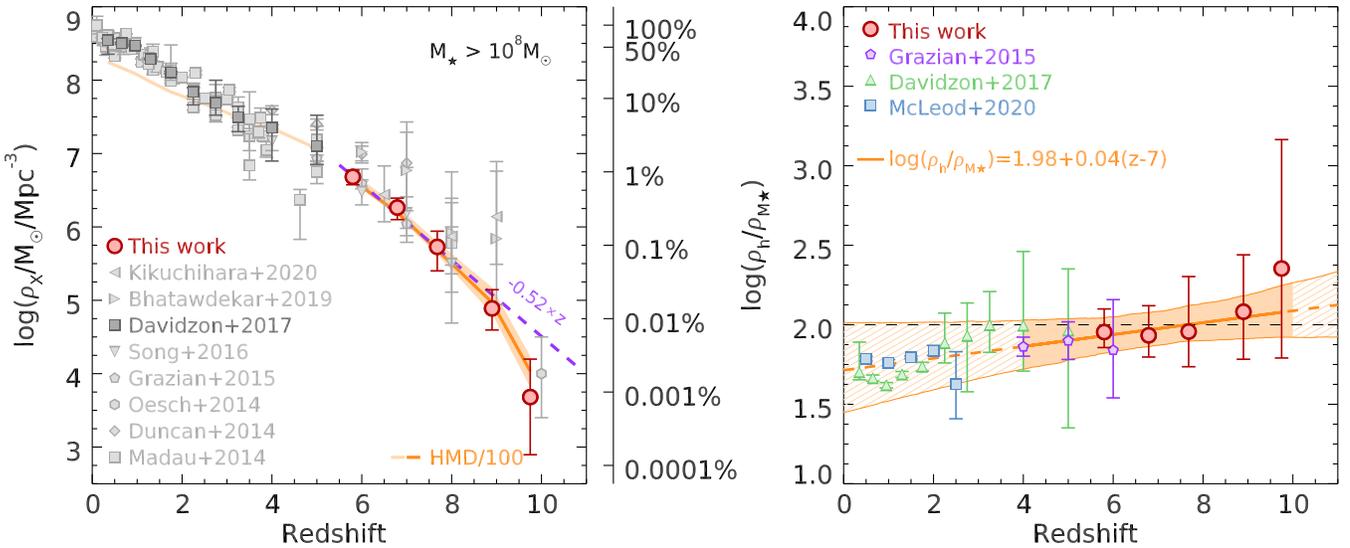}
\caption{\textbf{ [Left]:} Evolution of the stellar mass density (for galaxies with $\mathcal{M}_\star\ge 10^8 \mathcal{M}_\odot$) over $\sim13.5$~Gyr. The red filled circles correspond to our SMD values, with the purple dashed line being the fit to our values for redshifts less than $z\sim8$.  The grey symbols mark existing measurements as indicated by the legend, converted to the \citet{salpeter1955} IMF where necessary, and include the compilation of \citet{madau2014} - see text for details - and measurements from \citet{oesch2014, duncan2014, grazian2015, song2016,davidzon2017, bhatawdekar2019} and \citet{kikuchihara2020}. The orange curves correspond to the mass density of dark matter halos, rescaled by a factor $1/100$, recovered from abundance matching the SMF from this work at $z\sim6,7, 8, 9$ and $10$ (dark orange curve) and the SMF of \citet{davidzon2017} at $z<5$ (light orange curve). The additional vertical axis on the right indicates the fraction of the stellar mass density relative to that in the local Universe. \textbf{ [Right]:} Ratio between the Halo mass density and the stellar mass density, computed from our SMF measurements (filled red circles) and from the SMFs of \citet{grazian2015, davidzon2017} and \citet{mcleod2020}, as indicated by the legend. All HMD and SMD measurements were performed applying the same method (see Sections \ref{sect:SMD} and \ref{sect:HMD}). We excluded from our analysis the HMD at $z\sim3.25$  of \citet{mcleod2020} and at $z\sim7$ of \citet{grazian2015} given that those estimates are essentially undetermined in our analysis. The orange line and shaded region correspond to the linear fit to the $z\ge4$ measurements and $68\%$ confidence interval, respectively. These results suggest only a marginal evolution of the ratio $\rho_\mathrm{h}/\rho_\star\sim100$ from $z\sim10$ to $z\sim3-4$, and with minimal change even to $z\sim0$.\label{fig:SMD}}
\end{figure*}

We computed our stellar mass density (SMD) from $z\sim6$ to $z\sim10$ from the best-fit Schechter functions down to a consistent lower mass limit of $\mathcal{M}_\star=10^8\mathcal{M}_\odot$. The full set of measurements can be found in Table \ref{tab:SMD} and are presented in Figure \ref{fig:SMD}. 

In the same figure we also show recent estimates from the literature. Specifically, we include the compilation of \citet{madau2014} with the estimates of  \citet{li2009, gallazzi2008, moustakas2013, bielby2012,perez-gonzalez2008, ilbert2013, muzzin2013b, arnouts2007, pozzetti2010, kajisawa2009,marchesini2009,reddy2012, caputi2011, gonzalez2011,lee2012,yabe2009} and \citet{labbe2013}, and measurements from \citet{oesch2014, duncan2014, grazian2015, song2016,davidzon2017, bhatawdekar2019} and \citet{kikuchihara2020}. 

At $z\sim6$ and $z\sim7$ there is good consistency between the measurements from \citet{grazian2015,song2016} and \citet{kikuchihara2020} and our results, but those of \citet{duncan2014} and \citet{bhatawdekar2019} are somewhat higher, even though the large error bars at $z\sim7$ make all the results formally consistent. At $z\sim8$ the large uncertainties make essentially all the existing measurements consistent, ours, as well as those of \citet{labbe2013, song2016, bhatawdekar2019} and \citet{kikuchihara2020}, despite systematic differences of $\gtrsim0.5$\,dex. At $z\sim9$ the current estimates of \citet{bhatawdekar2019} and \citet{kikuchihara2020} lie above ours. This is not surprising, considering that our $z\sim9$ SMF is lower by $\sim1$\,dex than the corresponding SMF from those studies. Finally, at $z\sim10$ our measurement is $\sim0.35$\,dex lower than the previous measurement of \citet{oesch2014}, but the two are consistent at $1\sigma$.

Our results between $z\sim10$ and $z\sim6$ suggest a smooth evolution of the SMD, with an indication of more rapid evolution occurring in the first $\sim500-600$\,Myr of cosmic time up to $z\sim8$. Our SMD from $z\le 8$ to $z\sim6$ is consistent with an exponential increase with redshift, with a slope of $(-0.52\pm0.11)z$, steeper than the $\sim-0.28z$ observed at $z\sim0-3$ by \citet{mcleod2020} and marginally consistent ($1.5\sigma$) with the $-0.36z$ dependence found by \citet{dayal2018} for $4<z<10$. At $z\sim10$ our SMD value lies below the extrapolation of the relation that is seen from $z\le 8$ to later times, indicating a fast buildup of stellar masses at these very early epochs. Strikingly, only a tiny fraction ($\sim5\times10^{-6}$) of today's stellar mass was already in place at $z\sim10$.  By $z\sim6$ (i.e., when the Universe was only $\sim1$\,Gyr old, or just $\sim500$\,Myr after the epoch corresponding to $z\sim10$), the density had increased  by $\sim1000\times$ from $z\sim10$! The resulting stellar mass density was then $\sim1-2\%$ of today's value. The evolution that followed happened at a much slower pace after $z\sim6$, requiring $\sim13$Gyr ($\sim93\%$ of cosmic history) for the SMD to grow the final factor $\sim50-100$.

\subsection{Dark matter halos}

According to the concordance galaxy formation scenario (e.g., \citealt{rees1977, white1978, fall1980} -- see also \citealt{baugh2006} for a review), the assembly of stellar mass at early times is driven by the accretion of dark matter haloes, which, in turn, drive the accretion of cold gas onto the galaxy. The gas is finally converted into stars, modulo an efficiency, which in general can depend on the mass of the dark matter halo. In this section we leverage our SMF estimates to probe the relation between stellar mass and halo mass in the first $\sim1$\,Gyr of cosmic history.

\subsubsection{Halo-mass density}
\label{sect:HMD}

\begin{figure*}
\includegraphics[width=18cm]{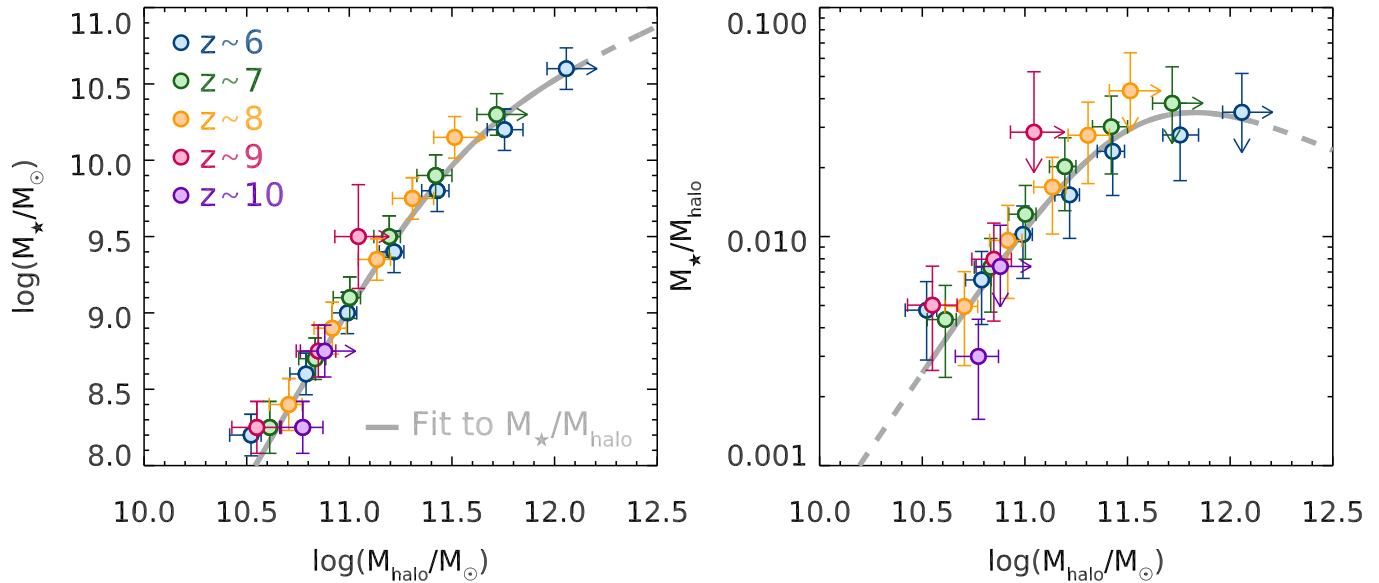}
\caption{{[\bf Left panel:]} Stellar mass as a function of the halo mass recovered from abundance matching the $V_\mathrm{max}$ estimates of the SMF presented in this work. Each color refers to a specific  redshift bin as labeled at the top-left corner. 	\textbf{ [Right panel:]} Ratio between the stellar mass and the halo mass, as a function of the halo mass, for the same redshift bins presented in the left-side panel. Our results are consistent with marginal evolution, or even no evolution at all, of the stellar-to-halo mass ratio between $z\sim10$ and $z\sim6$. The grey curve corresponds to our best-fit $\mathcal{M}_\star/\mathcal{M}_\mathrm{halo}$ parameterization (see Eq. \ref{eq:shmr}), fitted to all the redshift-merged estimates. The solid points with arrows mark those halo mass measurements whose lower uncertainty could not be determined because of the limited range available for $v_\mathrm{acc}$. The overlap of $\mathcal{M}_\star/\mathcal{M}_\mathrm{halo}$ across the redshift range of our study indicates that the star formation efficiency does not evolve in the first Gyr.  \label{fig:mh_mstar}}
\end{figure*}

\begin{deluxetable*}{cccccc}
\tablecaption{Dark matter halo masses and stellar-to-halo mass ratios \label{tab:SHMR}}
\tablehead{\colhead{Redshift} & \colhead{$\log(\mathcal{M}_\star/\mathcal{M}_\odot)$\tablenotemark{a}} & \colhead{$\log(\mathcal{M}_\mathrm{h}/\mathcal{M}_\odot)$\tablenotemark{b}} & \colhead{$\mathcal{M}_\star/\mathcal{M}_\mathrm{h}$\tablenotemark{c}}  & \colhead{$\log(\mathrm{cum.~den.}/\mathrm{Mpc}^{-3})$\tablenotemark{d}} & \colhead{$\log(v_\mathrm{acc}/\mathrm{km}/\mathrm{s}^{-1})$\tablenotemark{e}}
}
\startdata
$ 5.80$ & $ 8.20\pm0.14$ & $10.52^{+0.05}_{-0.10}$ & $ 4.8^{+1.6}_{-1.9} \times 10^{-3} $ & $-2.19^{+0.04}_{-0.04}$ & $ 2.01^{+0.01}_{-0.01}$ \\
        & $ 8.60\pm0.14$ & $10.79^{+0.05}_{-0.08}$ & $ 6.5^{+2.1}_{-2.3}  \times 10^{-3} $ & $-2.62^{+0.03}_{-0.03}$ & $ 2.10^{+0.01}_{-0.01}$ \\
        & $ 9.00\pm0.14$ & $10.99^{+0.05}_{-0.07}$ & $ 10.^{ +3.}_{ -4.}  \times 10^{-3} $ & $-2.97^{+0.05}_{-0.05}$ & $ 2.16^{+0.01}_{-0.01}$ \\
        & $ 9.40\pm0.14$ & $11.22^{+0.05}_{-0.07}$ & $ 15.^{ +5.}_{ -5.}  \times 10^{-3} $ & $-3.40^{+0.05}_{-0.05}$ & $ 2.24^{+0.01}_{-0.01}$ \\
        & $ 9.80\pm0.14$ & $11.43^{+0.06}_{-0.08}$ & $ 24.^{ +8.}_{ -8.}  \times 10^{-3} $ & $-3.85^{+0.12}_{-0.13}$ & $ 2.31^{+0.02}_{-0.02}$ \\
        & $10.20\pm0.14$ & $11.76^{+0.09}_{-0.08}$ & $ 28.^{+10.}_{-10.}  \times 10^{-3} $ & $-4.65^{+0.22}_{-0.31}$ & $ 2.43^{+0.04}_{-0.03}$ \\
        & $10.60\pm0.14$ & $>11.97$ & $ <52  \times 10^{-3} $ & $<-5.17$ & $ >2.50$ \\
& & & & & \\
$ 6.79$ & $ 8.25\pm0.17$ & $10.61^{+0.05}_{-0.09}$ & $ 4.3^{+1.8}_{-1.9} \times 10^{-3} $ & $-2.69^{+0.06}_{-0.06}$ & $ 2.07^{+0.01}_{-0.01}$ \\
        & $ 8.70\pm0.14$ & $10.83^{+0.05}_{-0.08}$ & $ 7.4^{+2.5}_{-2.7}  \times 10^{-3} $ & $-3.11^{+0.05}_{-0.05}$ & $ 2.14^{+0.01}_{-0.01}$ \\
        & $ 9.10\pm0.14$ & $11.00^{+0.05}_{-0.08}$ & $ 13.^{ +4.}_{ -5.}  \times 10^{-3} $ & $-3.45^{+0.07}_{-0.07}$ & $ 2.20^{+0.01}_{-0.01}$ \\
        & $ 9.50\pm0.14$ & $11.19^{+0.05}_{-0.08}$ & $ 20.^{ +7.}_{ -7.}  \times 10^{-3} $ & $-3.87^{+0.09}_{-0.09}$ & $ 2.27^{+0.01}_{-0.01}$ \\
        & $ 9.90\pm0.14$ & $11.42^{+0.08}_{-0.09}$ & $ 30.^{+11.}_{-11.}  \times 10^{-3} $ & $-4.42^{+0.19}_{-0.25}$ & $ 2.34^{+0.04}_{-0.02}$ \\
        & $10.30\pm0.14$ & $>11.62$ & $ <55  \times 10^{-3} $ & $<-4.93$ & $ >2.41$ \\
& & & & & \\
$ 7.68$ & $ 8.40\pm0.17$ & $10.71^{+0.07}_{-0.10}$ & $ 5.0^{+2.1}_{-2.2} \times 10^{-3} $ & $-3.33^{+0.10}_{-0.10}$ & $ 2.13^{+0.02}_{-0.02}$ \\
        & $ 8.90\pm0.17$ & $10.92^{+0.07}_{-0.09}$ & $ 9.6^{+4.1}_{-4.2}  \times 10^{-3} $ & $-3.80^{+0.13}_{-0.14}$ & $ 2.20^{+0.02}_{-0.02}$ \\
        & $ 9.35\pm0.14$ & $11.13^{+0.07}_{-0.09}$ & $ 16.^{ +6.}_{ -6.}  \times 10^{-3} $ & $-4.31^{+0.16}_{-0.16}$ & $ 2.27^{+0.02}_{-0.02}$ \\
        & $ 9.75\pm0.14$ & $11.31^{+0.10}_{-0.10}$ & $ 28.^{+11.}_{-11.}  \times 10^{-3} $ & $-4.83^{+0.29}_{-0.44}$ & $ 2.34^{+0.05}_{-0.04}$ \\
        & $10.15\pm0.14$ & $>11.41$ & $ <63  \times 10^{-3} $ & $<-5.10$ & $ >2.37$ \\
& & & & & \\
$ 8.90$ & $ 8.25\pm0.17$ & $10.55^{+0.12}_{-0.12}$ & $ 5.0^{+2.4}_{-2.4} \times 10^{-3} $ & $-3.87^{+0.31}_{-0.45}$ & $ 2.13^{+0.06}_{-0.05}$ \\
        & $ 8.75\pm0.17$ & $10.85^{+0.09}_{-0.11}$ & $ 8.0^{+3.5}_{-3.7}  \times 10^{-3} $ & $-4.58^{+0.22}_{-0.27}$ & $ 2.23^{+0.03}_{-0.03}$ \\
        & $ 9.50\pm0.34$ & $>10.93$ & $ <55  \times 10^{-3} $ & $<-4.86$ & $ >2.26$ \\
& & & & & \\
$ 9.75$ & $ 8.25\pm0.17$ & $10.77^{+0.10}_{-0.11}$ & $ 3.0^{+1.4}_{-1.4} \times 10^{-3} $ & $-4.96^{+0.34}_{-0.46}$ & $ 2.22^{+0.05}_{-0.04}$ \\
        & $ 8.75\pm0.17$ & $>10.76$ & $ <11.2  \times 10^{-3} $ & $<-4.91$ & $ >2.22$ \\
\enddata
\tablecomments{Upper/lower limits are $2\sigma$.}
\tablenotetext{a}{Central value of the stellar mass bin adopted for our SMF estimates. The uncertainty corresponds to $\pm34\%$ the width of the mass bin.}
\tablenotetext{b}{Halo mass recovered through our abundance matching procedure.}
\tablenotetext{c}{Stellar-to-halo mass ratio. Uncertainties correspond to the propagation of the uncertainties in the stellar and halo masses.}
\tablenotetext{d}{Cumulative density adopted for the abundance matching, computed for the central mass of each stellar mass bin.} 
\tablenotetext{e}{Accretion velocity of the dark matter halos for the cumulative density in the adjacent column (d).}
\end{deluxetable*}

We applied abundance matching techniques (\citealt{kravtsov2004,tasitsiomi2004,vale2004,conroy2006})  to our SMF estimates to evaluate whether the rapid growth of the stellar mass density observed in Figure \ref{fig:SMD} is matched by that of the dark matter halos. 

Dark matter halos can undergo more significant stripping before being accreted than their baryonic counterparts (\citealt{conroy2006, trujillo-gomez2011, reddick2013, wechsler2018, campbell2018}). Given this, recent work suggests that the peak maximum velocity of the particles in the dark matter halo across its formation history (commonly denoted as $V_\mathrm{peak}$), or the maximum circular velocity of a halo at the time of accretion ($v_\mathrm{acc}$), constitute a better match to the baryonic properties of galaxies  and can better reproduce the two-point correlation function (e.g., \citealt{conroy2006, reddick2013}). For our exercise, we therefore matched the cumulative densities of the SMF to $\mathcal{M}_\star=10^8\mathcal{M}_\odot$ to those of $v_\mathrm{acc}$ provided by the Bolshoi N-body numerical simulation (\citealt{klypin2016}). In particular, we adopted the Bolshoi-Planck run, based on a \citet{planck2014_cosmology} $\Lambda$CDM cosmology with parameters $h = 0.677$, $\Omega_m = 0.307$, $\Omega_\Lambda = 0.693$, $n_s = 0.96$ and $\sigma_8 = 0.823$\footnote{The cosmological parameters adopted for the Bolshoi simulation differ from our $0.7,0.3,0.7$ fiducial cosmology. These differences systematically affect the estimates of volume densities and stellar masses, and could therefore potentially affect our abundance matching analysis. Adopting $h=0.67$ instead of $h=0.7$ would result in stellar masses larger by $\sim0.03$ dex. The corresponding shift of the SMFs would mimick an increase of the volume densities ($\sim+0.054$ dex for $\alpha=-1.8$). However, for $h=0.67$ the volume densities would be smaller by $\sim0.04$ dex, mitigating most of the apparent increase in volume density resulting from the higher stellar masses. The differences in $\Omega_\Lambda$ and $\Omega_m$ result in even smaller corrections. Such very small residual differences allow us to conclude that our abundance matching results are robust against the marginal differences between the cosmological parameters used for our SMF estimates and those adopted for the Bolshoi simulation.}.  The simulation was run in a box of $250h^{-1}$\,Mpc side, and includes $2048^3$ particles, allowing us to resolve halos with a mass of $10^{10}\mathcal{M}_\odot$ (see \citealt{klypin2016} and \citealt{rodriguez-puebla2016} for details on the simulation). Dark matter halo masses and $68\%$ confidence intervals were then obtained as the median and $16$th- and $84$th-percentiles of the halo masses included in the range of the recovered $v_\mathrm{acc}$. The results of this procedure are reported in Table \ref{tab:SMD}.  We also repeated the abundance matching procedure adopting the halo mass functions of \citet{behroozi2013} generated by the \textsc{HMFcalc} tool \citep{murray2013} and found that the halo masses differ by $\lesssim0.04$\,dex from those computed with the $v_\mathrm{acc}$ abundance, increasing our confidence in the results.

While the halo mass densities are $\sim2$ orders of magnitude higher than the corresponding stellar mass densities, it is quite informative to compare the \textit{rate} at which they grow to that of the stellar mass density. Therefore in Figure \ref{fig:SMD} (Left) we plot the halo mass densities after rescaling them by a factor $0.01$. It is striking that, despite the large uncertainties, particularly at the highest redshifts, the growth of the stellar mass density follows that of the halo mass; in particular between $z\sim6$ and $z\sim9$ there seems to be an almost $1:1$ relation between the two rates. We repeated the same procedure at lower redshifts adopting the SMF of \citet{davidzon2017}. The resulting HMD estimates, after being rescaled by the same $0.01$ factor, are marked in Figure \ref{fig:SMD} (Left) with the lighter orange curve. Because of the limited depth available for the SMFs of \citet{davidzon2017}, we compare, in the right panel of Figure \ref{fig:SMD}, our measurements of the ratio of the stellar mass density and the dark matter halo density to those we obtain with the same procedure using the SMFs of \citet{grazian2015, davidzon2017} and \citet{mcleod2020}. These measurements show that the rate of growth of the stellar mass assembly is still very similar to that of the dark matter halos down to $z\sim4$. The marginal evolution is confirmed by a linear fit to the logarithm of the ratio between the two densities at $z\ge4$, resulting in:
\begin{equation}
\log(\rho_\mathrm{h}/\rho_\mathcal{M_\star})=(1.976\pm0.104)+(0.037\pm0.041)\times(z-7)
\end{equation}

Remarkably, our analysis shows that the stellar and halo mass densities show a consistent trend in their ratio to $z\sim0$, with both having increased by $5$ orders of magnitude between $z\sim10$ and $z\sim0$. Nonetheless, their ratio in Figure \ref{fig:SMD} (right) has changed by a strikingly small $0.3-0.5$\,dex over this same redshift range ($\sim96\%$ of cosmic history), especially when compared to the $\sim5$ dex growth in both $\rho_\mathrm{h}$  and $\rho_{\star}$.

\begin{figure*}
\includegraphics[width=18cm]{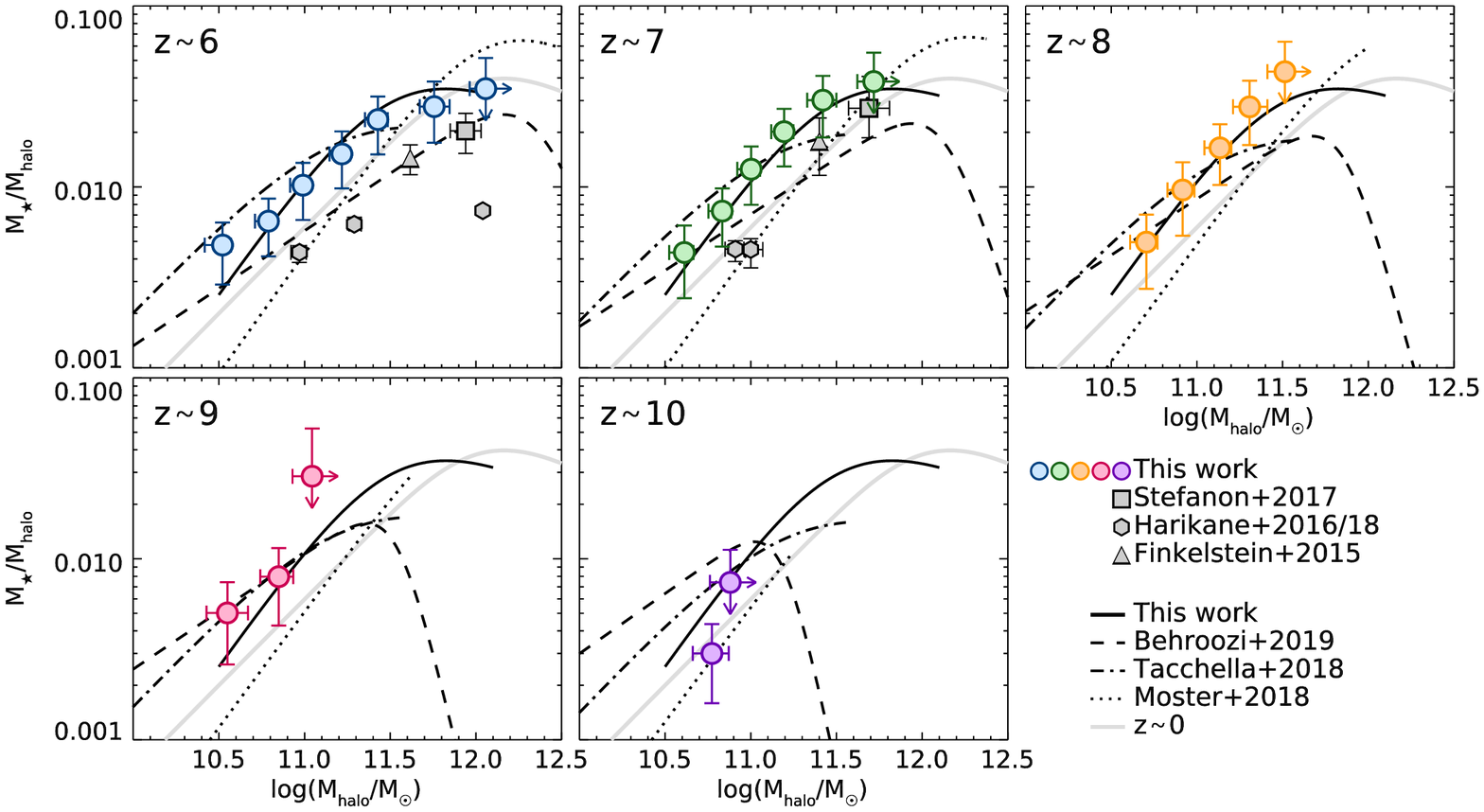}
\caption{Comparison of the stellar-to-halo mass ratios estimated in this study to previous observational estimates of \citet{finkelstein2015b, harikane2016, harikane2018} and \citet{stefanon2017b} (symbols as per the bottom-right legend). Also presented is our redshift-independent best-fitting relation, and the redshift-evolving stellar-to-halo mass relations from the semi-analytic model of \citet{behroozi2019}, the moderately evolving SHMR of \citet{moster2018} and the marginally evolving SHMR from the semi-analytic models of \citet{tacchella2018} (legend at the bottom right). The grey solid curve corresponds to the $z\sim0$ SHMR from \citet{behroozi2013} that we display as reference. We used the relations derived in Appendix~\ref{sect:app_cosmology} to convert the halo mass estimates of \citet{finkelstein2015b} and the curves of \citet{tacchella2018}  from a WMAP7 cosmology into one consistent with that of \citet{planck2016_cosmology}. The solid points with arrows mark those halo mass measurements whose  lower uncertainty remained undefined because of the limited range available for $v_\mathrm{acc}$. The broad agreement with the predictions with constant SHMR (see text) provide further support to a non-evolving SHMR in the early universe. \label{fig:shmr_models}}
\end{figure*}

The above results are qualitatively consistent with the co-evolution between the star-formation rate density and the halo mass accretion rate recently found by \citet{oesch2018}, indicating a scenario where the efficiency of star formation remained approximately constant through the first $\sim1.5$\,Gyr of cosmic history, as also suggested by some of the recent models (e.g., \citealt{mason2015, mashian2016, wilkins2017, tacchella2018, bhowmick2018, ma2018, park2019, bouwens2021,hutter2021}) and observations (e.g., \citealt{durkalec2015, stefanon2017b, harikane2018}). These results are also qualitatively consistent with the co-evolution between the specific SFR and the specific dark matter halo mass accretion rate found in recent studies (e.g., \citealt{stefanon2021c}).

\subsubsection{Stellar-to-Halo mass ratio}
\label{sect:SHMR}

The ratio between the stellar and the halo mass is a proxy for the efficiency of the conversion of cold gas into stars. The co-evolution between dark-matter halos and stellar mass presented in the previous section is dominated by galaxies with $\mathcal{M}_\star \sim 10^8\mathcal{M}_\odot$ because of their larger volume densities. In this section we explore in more detail the relation between the stellar mass and halo mass across the range of stellar masses probed in our SMF estimates.

We adopted the abundance matching tools discussed in the previous section to estimate the dark matter halo masses for each specific stellar mass bin. However, for this analysis we are not constrained to match the same limit in stellar mass across the different redshifts, as instead it was the case for the SMD discussed in Section \ref{sect:SMD}. We therefore computed the reference cumulative densities by numerically integrating our $V_\mathrm{max}$ measurements in correspondence of each value in stellar mass. For this, we adopted the center of the stellar mass bin as reference value, reducing by $50\%$ the amplitude of the lowest stellar mass bin in each computation. Our results do not significantly differ when we use the Schechter parameterizations instead of the $v_\mathrm{max}$ estimates. We set the uncertainties in stellar mass to $68\%$ ($\pm34\%$) of the width of the corresponding stellar mass bin, assuming an approximately uniform distribution of stellar mass within each stellar mass bin. The resulting halo masses and stellar-to-halo mass ratios (SHMR) are listed in Table \ref{tab:SHMR}, and are graphically presented in Figure \ref{fig:mh_mstar}.

Our measurements indicate a monotonic increase at all redshifts, as expected from the extrapolation of results at lower redshifts (see e.g., \citealt{wechsler2018, legrand2019, girelli2020} and references therein). Remarkably, the  $(\mathcal{M}_\star,  \mathcal{M}_\mathrm{halo}$) pairs computed for each redshift $z\sim6, 7, 8, 9,$ and $10$ overlap within the nominal uncertainties over most of the range in halo masses. This further supports our result of an essentially constant efficiency of star formation at these epochs. Nonetheless, the $z\sim9$ and $z\sim10$ estimates have a lower significance ($\sim 1-2\sigma$), potentially hiding any evolution in the first $\sim600$Myr.\\

Because our $(\mathcal{M}_\star, \mathcal{M}_\mathrm{halo})$ measurements do not strongly depend on redshift, we fitted the following parametric form  (\citealt{moster2010} - see also \citealt{yang2003}) after merging all sets of measurements: 
\begin{equation}
\frac{\mathcal{M}_\star}{\mathcal{M}_\mathrm{halo}}=2N\left[ \left(\frac{\mathcal{M}_\mathrm{halo}}{\mathcal{M}_c}\right)^{-\beta} + \left(\frac{\mathcal{M}_\mathrm{halo}}{\mathcal{M}_c}\right)^{\gamma} \right]^{-1}
\label{eq:shmr}
\end{equation}
Here $N$ is a normalization factor, $\mathcal{M}_{c}$ is a characteristic halo mass where the star-formation efficiency is maximized, while $\beta$ and $\gamma$ are the slopes of the low-mass and high-mass regimes, respectively. Given that  our range in mass does not probe masses larger than $\approx \mathcal{M}_c$ needed to constrain $\gamma$, we assumed $\gamma=0.4$ (\citealt{tacchella2018}). For this same reason, our constraints on $\mathcal{M}_c$ should be taken with caution. Our fit results in $\beta=1.35\pm0.26$, $\log\mathcal{M}_c/\mathcal{M}_\odot=11.5\pm0.2$ and $N=0.0297\pm0.0065$. The corresponding parameterization is presented in Figure \ref{fig:mh_mstar} with the solid grey curve.

In Figure \ref{fig:shmr_models} we compare our estimates with existing determinations. Specifically, at $z\sim6$ and $z\sim7$ we included the measurements of \citet{finkelstein2015b} which are based on abundance matching the UV LF, the estimates of \citet{harikane2016,harikane2018}, which rely on the two-point correlation function of Lyman Break galaxies, and those of \citet{stefanon2017b} obtained applying the abundance matching to the rest-frame optical LF. The estimates of \citet{finkelstein2015b} and \citet{tacchella2018} assumed a WMAP7 cosmology \citep{komatsu2011} shifting the halo masses towards higher values. Given this we converted them to a \citet{planck2016_cosmology} cosmology applying the correction described in Appendix \ref{sect:app_cosmology}. Furthermore, we multiply by a factor 1.7 the stellar-to-halo mass values of \citet{harikane2016,harikane2018} to convert them from a \cite{chabrier2003} to a \citet{salpeter1955} IMF. To our knowledge, there are no other estimates to date of the stellar-to-halo mass ratios at $z\sim8$, $z\sim9$ and $z\sim10$. 

Our estimates are consistent with those of \citet{stefanon2017b} at $<1\sigma$ and of \citet{finkelstein2015b} at $\gtrsim1\sigma$, both of which are based on abundance matching techniques; however, they are factor $3-4\times$ higher (corresponding to a $\sim3\sigma$ difference) than those of \citet{harikane2016,harikane2018}, which were derived from clustering measurements. In the same panels we also present SHMR from three recent models which use different assumptions on the evolution of the SHMR with cosmic time: \citet{tacchella2018} assumed the SHMR to be approximately constant above $z\sim4$; \citet{moster2018} linked the star-formation rate to the halo accretion rate through a redshift-dependent parametric baryon conversion efficiency; finally, \citet{behroozi2019} did not introduce any correlation between the evolution of the dark matter halos and (baryonic) galaxy assembly, finding a SHMR increasing with redshift  (see also \citealt{behroozi2013}, but see \citealt{zhu2020}). Figure \ref{fig:shmr_models} shows that our measurements are  generally in good agreement with the predictions of \citet{tacchella2018} over the full redshift range probed here, and with those of \citet{behroozi2019} at $z\sim8-10$, further supporting a non-evolving SHMR in the early Universe.

\section{Summary and Conclusions}
\label{sect:conclusions}

The new deep, wide GREATS IRAC dataset, combined with a much larger \textit{Hubble} sample, has allowed us to derive statistically-robust stellar mass functions (SMF) from $\sim 800$ galaxies at redshifts between $z\sim6$ and $z\sim10$.  The comprehensive catalog of Lyman-break galaxies (LBG) was assembled from the source lists of \citet{bouwens2015, bouwens2016, bouwens2019a} and \citet{oesch2018} over the GOODS, the HUDF/XDF and all five CANDELS fields. Our stellar mass samples are distinctive compared to previous studies at similar redshifts due to our use (1) a much deeper wide area ($\sim200$\,hour) \textit{Spitzer}/IRAC imaging dataset at  $3.6\mu$m and $4.5\mu$m from the GOODS Re-ionization Era wide Area Treasury from Spitzer (GREATS) program (PI: I. Labb\'e - \citealt{stefanon2021a}) and (2) a $3\times$ larger search volume than previous HST-based galaxy SMFs. These new deep \textit{Spitzer} data greatly increased the number, and fraction, of IRAC-detected sources from the UV catalogs. For example, $>50\%$ of the sources with stellar masses $\mathcal{M}_\star>10^{8}\mathcal{M}_\odot$ showed  $\ge 2\sigma$ detection in the rest-frame optical. Constraining the UV-selected sources with a large fraction of IRAC measurements significantly increased the robustness of our stellar mass measurements. 

Our SMFs were derived using the $V_\mathrm{max}$ method of \citet{avni1980} on individual sources. Schechter fits to the $z\sim6-8$ SMFs suggest a non-evolving low-mass end slope $\alpha\sim-1.8$, broadly consistent with previous estimates. The $\chi^2$ contours indicate the SMFs evolve in both the characteristic stellar mass $\mathcal{M}^*$ and the number density normalization factor  $\phi^*$. The stellar mass density (SMD) increases by $\sim1000 \times$ in the 0.5 Gyr between $z\sim10$ and $z\sim6$, with an evolution qualitatively consistent with that of the star-formation rate density (see, e.g., \citealt{oesch2018}). This rapid growth during the first Gyr in just 500 Myr since $z\sim10$ contrasts with the further, and slower, $\sim100 \times$ increase over the next ~13 Gyr to $z\sim0$.

We performed abundance matching of our SMFs to the Bolshoi-Planck simulation (\citealt{klypin2016}). Our analysis shows that the SMD follows the growth rate of the halo mass density from $z\sim10$ to $z\sim3-4$.  In particular, a fit to the ratio between the dark-matter halo mass density and the stellar mass density at $z\ge4$ gives:
\begin{equation}
\log(\rho_\mathrm{h}/\rho_\mathcal{M_\star})=(1.976\pm0.104)+(0.037\pm0.041)\times(z-7)
\end{equation}

\noindent  Remarkably, we find no evidence for evolution in the stellar-to-halo mass ratios from $z\sim10$ to $z\sim6$  for galaxies in the $10^8<\mathcal{M}_\star/\mathcal{M}_\odot\lesssim10^{10}$ stellar mass range. This is even more remarkable given the three orders-of-magnitude increase in the SMD in the 500 Myr from $z\sim10$ to $z\sim6$ noted above. Our results at the earliest times fit well with those found previously for later times $z\le6$.  Our results furthermore indicate at most a marginal evolution of the star-formation efficiency at these early epochs, nicely consistent with many recent empirical models (e.g., \citealt{tacchella2018, park2019}).  \\

In the near future, the \textit{James Webb Space Telescope (JWST)} will significantly increase the sensitivity of the flux measurements at $3-5\mu$m.  Source confusion, which can be challenging to overcome with the \textit{Spitzer}/IRAC data, will be much less of a concern due to an impressive $\sim10 - 15\times$ reduction in the PSF FWHM at $3-5\mu$m.  The substantially improved flux measurements at longer wavelengths will also come with improvements in the efficiency with which sources are selected.  

While our new \textit{Hubble} and \textit{Spitzer} SMF results have yielded striking insights into the lack of significant changes in the stellar-to-halo mass ratios, and in the star-formation efficiency, in the first Gyr from $z\sim10$ to $z\sim6$ when truly dramatic growth is occurring in the SMD, \textit{JWST} is poised to take us even further. \textit{JWST} will provide more detailed insights and verification at $z\le10$, but, crucially, will reveal what happens to the star-formation efficiency prior to $z\sim10$, into the epoch of the "first galaxies" during the first $500$ Myr of cosmic time.

\appendix

\section{Completeness estimate}
\label{app:LF_cc}

\begin{figure*}
\includegraphics[width=18cm]{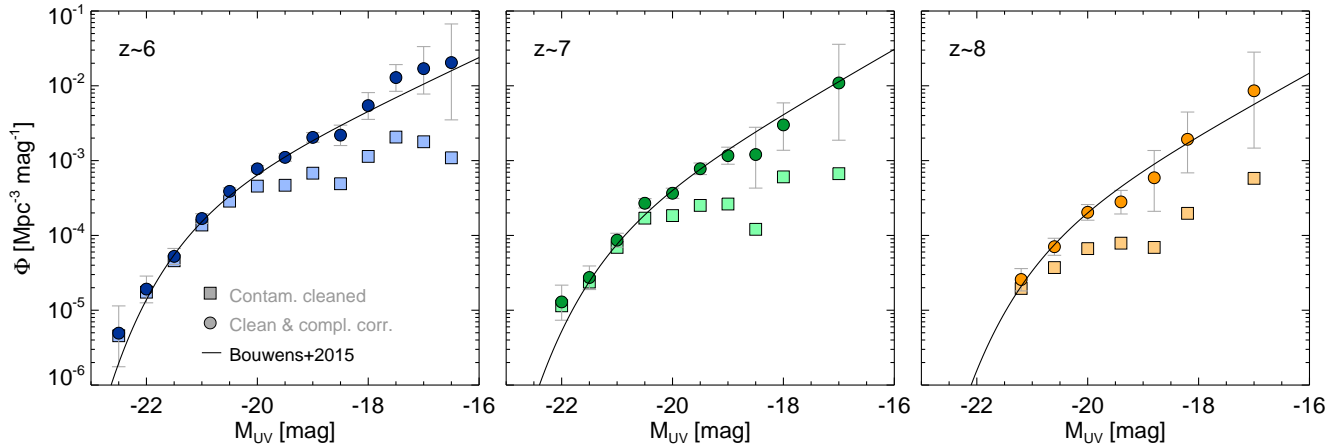}
\caption{Effectiveness of recovering the HST UV luminosity distribution after excluding sources from the HST sample with $>65\%$  contamination by neighbours in the IRAC $3.6\mu$m and $4.5\mu$m bands. Each panel refers to a specific redshift bin, as labelled in the top-left corner. The filled square symbols represent the UV luminosity function computed after cleaning the sample from the contaminated sources, while the filled circles mark the  volume density obtained after applying the completeness corrections computed through our Monte Carlo simulation (see Sect. \ref{sect:sample} for details). The solid black curve corresponds to the Schechter parameterization of \citet{bouwens2015}, adopted as reference. The small IRAC flux densities of faint sources imply lower contributions from the neighbouring objects are sufficient to satisfy the contamination criteria, making completeness corrections increasingly large for faint sources. Our completeness estimates allow us to recover the UV LF very well to $M_\mathrm{UV}\gtrsim -17.0$. However, the median corrections are $>10\times$ for $M_\mathrm{UV}\sim -16.75, -17.25$ and $\sim -17.5$ at $z\sim6, 7$ and $\sim8$, respectively,  making the corresponding volume densities at low luminosities highly uncertain. \label{fig:LF_cc}}
\end{figure*}

We assessed the statistical effects of our IRAC selection on our sample through a Monte Carlo simulation. This consisted in adding synthetic sources to the GREATS $3.6\mu$m and $4.5\mu$m mosaics and recovering their flux density and contamination using \textsc{Mophongo}. The synthetic sources were added at random locations across the $3.6\mu$m-band mosaic, and then at the same locations when adding sources in the $4.5\mu$m band. This procedure was repeated over a suitable range of flux densities. We estimated the fraction of sources excluded because of a high neighbour contamination by applying the same selection criteria adopted for the main sample assuming a flat $f_\nu$ SED.  To evaluate the effectiveness of the estimated correction, we computed the UV luminosity function from the HST sources free of those with neighbour contamination in the IRAC bands at $z\sim6,7$ and $8$ using the V$_\mathrm{max}$ formalism (\citealt{avni1980}) and weighting the volumes associated to each source by the estimated correction. Figure \ref{fig:LF_cc} presents the result of this exercise. Our completeness corrections are clearly larger for fainter sources in our samples. This is a result of the fact that for fainter sources, even a small contribution of light at $3.6\mu$m and $4.5\mu$m from their neighbors is sufficient to satisfy the $65\%$ IRAC contamination threshold for exclusion from our sample. Comparison of our UV LFs to those of \citet{bouwens2015} show a good agreement over the full range of luminosities and for all redshifts. However, the median of the corrections become very large ($>10\times$) for the faintest sources  $M_\mathrm{UV}\gtrsim -16.75, -17.25$ and $\sim -17.5$\,mag in our $z\sim6, 7$ and $\sim8$ selections, respectively, making the associated volume densities more uncertain. For this reason, in our analysis we exclude those measurements potentially affected by this aspect.

\section{Comparison of stellar mass estimates from our different methods}
\label{app:mstar_comp}

\begin{figure*}
\includegraphics[width=15cm]{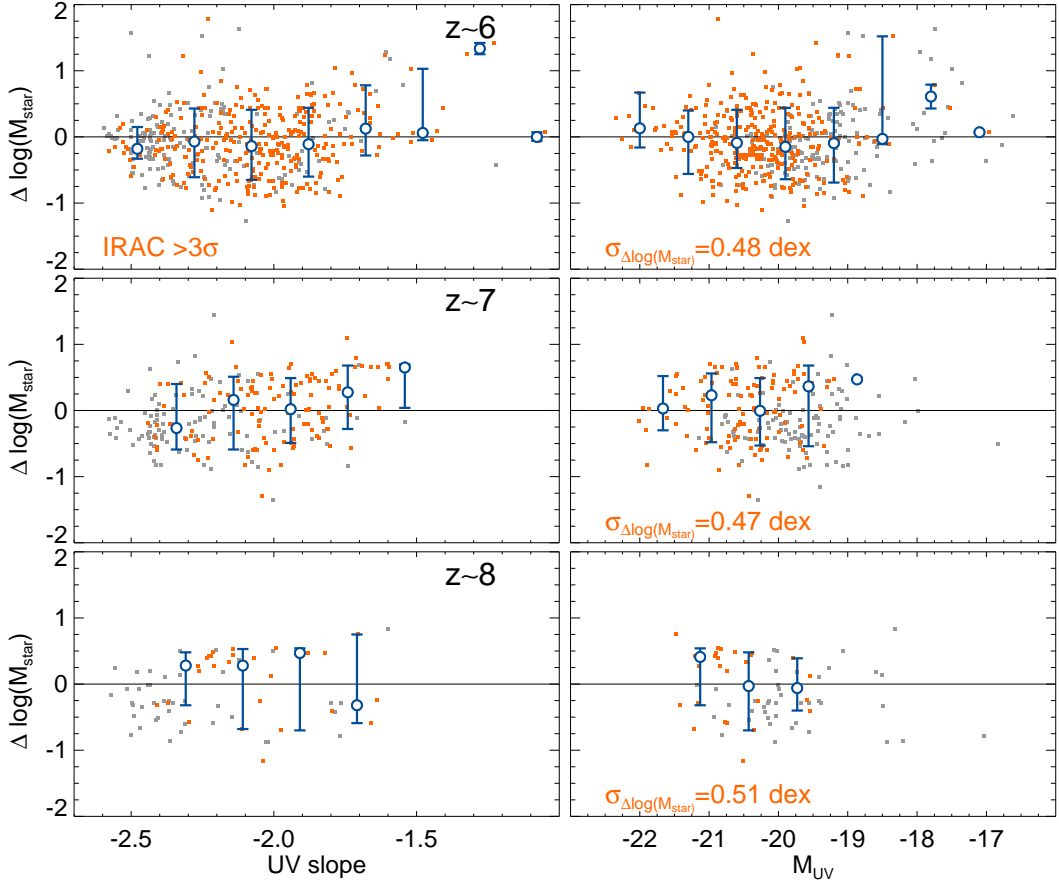}
\caption{Comparison between the stellar mass measurements obtained with SED templates that included emission by nebular continuum and lines ($\equiv M_\mathrm{star,full}$), and those where the rest-frame optical free from emission line contribution is reconstructed using the relation between the $H-[3.6]$ color and the UV slope ($\equiv M_\mathrm{star,H36}$) of \citet{stefanon2021c}. Top to bottom, the panels refer to the $z\sim6$, $z\sim7$ and $z\sim8$ redshift bins. For each redshift bin, the left panel presents $\Delta \log (M_\mathrm{star}) \equiv \log(M_\mathrm{star,full})-\log(M_\mathrm{star,H36})$ as a function of the UV slope, while the panels on the right present $\Delta \log(M_\mathrm{star})$ as a function of the UV luminosity $M_\mathrm{UV}$. Grey points correspond to the full sample, while orange points mark those sources with higher  significance in the two IRAC bands, as indicated by the label in the top-left panel. Finally, the blue points with error bars correspond to the median and 68\% confidence interval of the sources with higher IRAC significance. The relative consistency between our stellar mass estimates made using the full SED information for IRAC-detected sources and estimates made relying on the UV slope information alone (exploiting a relationship we found in \citealt{stefanon2021c}) gives us confidence in using this prescription for sources lacking clear $2\sigma$ detections with IRAC. \label{fig:delta_Mstar}}
\end{figure*}

In Figure \ref{fig:delta_Mstar} we compare the stellar mass estimates of the full sample at $z\sim6-8$ estimated using the methods described in Section \ref{sect:stellar_pop}, \ref{sect:mstar_det} and \ref{sect:mstar_undet}, as a function of the UV slope and UV luminosity, in the three redshift bins. The stellar mass estimates from the updated IRAC bands for sources detected at $>3\sigma$ significance in both IRAC bands are on average consistent with those obtained with the SED analysis on the original photometry. Because the systematic differences are marginal and within the $1\sigma$ dispersions (see Figure \ref{fig:delta_Mstar}), we concluded that the new set of stellar masses can confidently be used for those sources with $<2\sigma$ significance in at least one of the two IRAC bands for the $z\sim6, 7$ and $z\sim8$ samples.

\section{Stellar mass-to-light ratios}
\label{app:mstar-muv}

\begin{figure*}
\includegraphics[width=18.5cm]{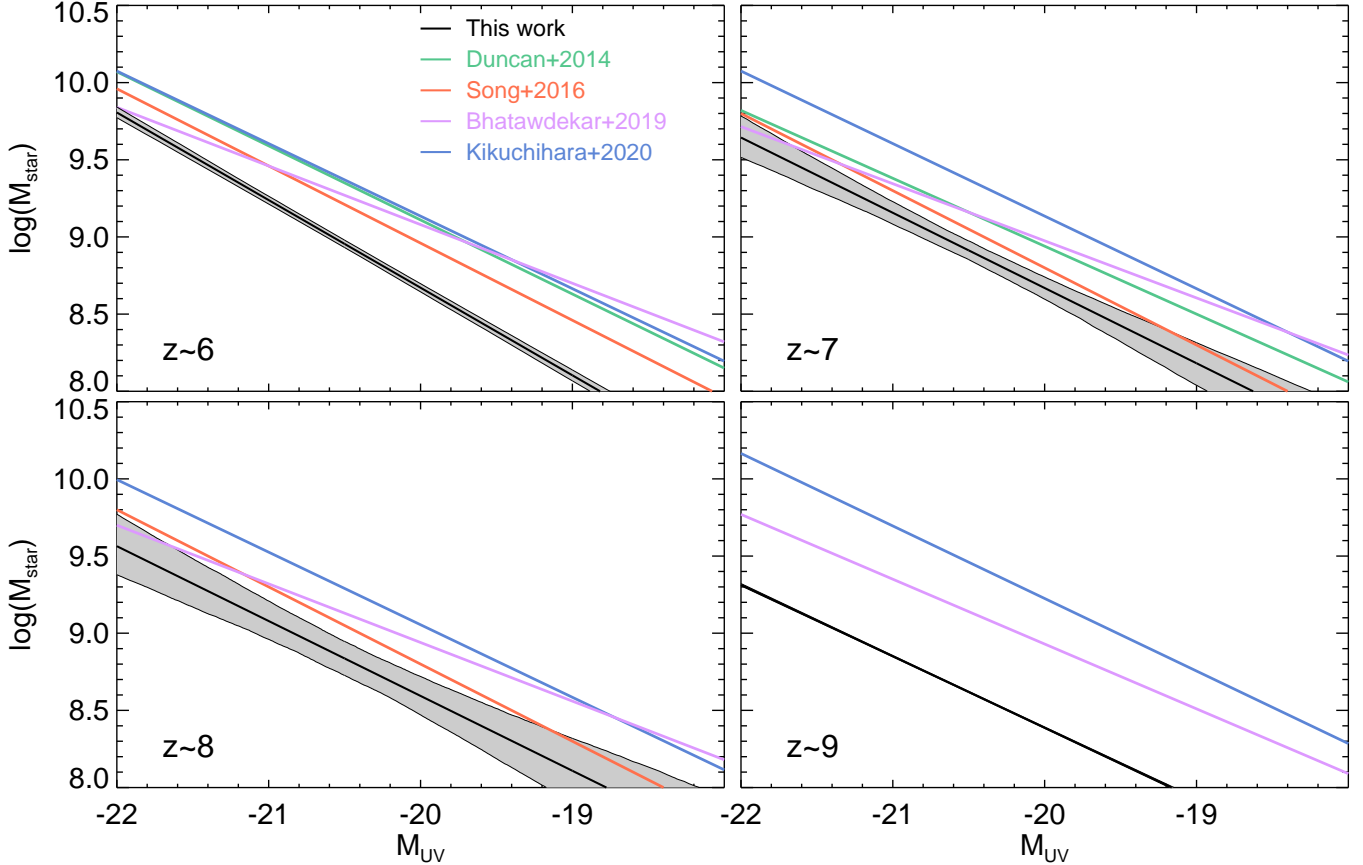}
\caption{Comparison of our $\mathcal{M}_\star-M_\mathrm{UV}$ relation to some of the most recent estimates in the literature, as indicated by the legend in the top-left panel. The grey-shaded areas encompass the $68\%$ confidence interval from our best-fit values. The tight relationship at $z\sim9$ is likely a consequence of similar SEDs resulting from our procedure of reconstructing the flux densities in the IRAC bands discussed in Section \ref{sect:mstar_z9}.
  \label{fig:comp_mstar-muv}}
\end{figure*}

In Figure \ref{fig:comp_mstar-muv} we compare our estimates of the $\mathcal{M}_\star-M_\mathrm{UV}$ relationship at $z\sim6,7,8$ and $9$ to those at similar redshifts from recent determinations (\citealt{duncan2014, song2016, bhatawdekar2019} and \citealt{kikuchihara2020}). Our recovered slopes are in general consistent with previous determinations. Interestingly, at $z\sim6,7$ and $8$ our stellar masses seem to be $\sim0.2-0.3$\,dex lower than the average of previous estimates. At $z\sim9$ this difference increases to $\sim0.5-1.0$\,dex. This difference could, at least in part, be due to the strong emission lines recovered from our improved IRAC colors which have higher S/N than in previous studies, implying younger stellar populations and lower $\mathcal{M}_\star/L_\mathrm{UV}$ ratios.  The use of the higher S/N GREATS data suggests that our results are likely to be representative of the true values.

\section{Conversion of SHMR based on WMAP7 cosmology into Planck cosmology}
\begin{figure*}
\includegraphics[width=18cm]{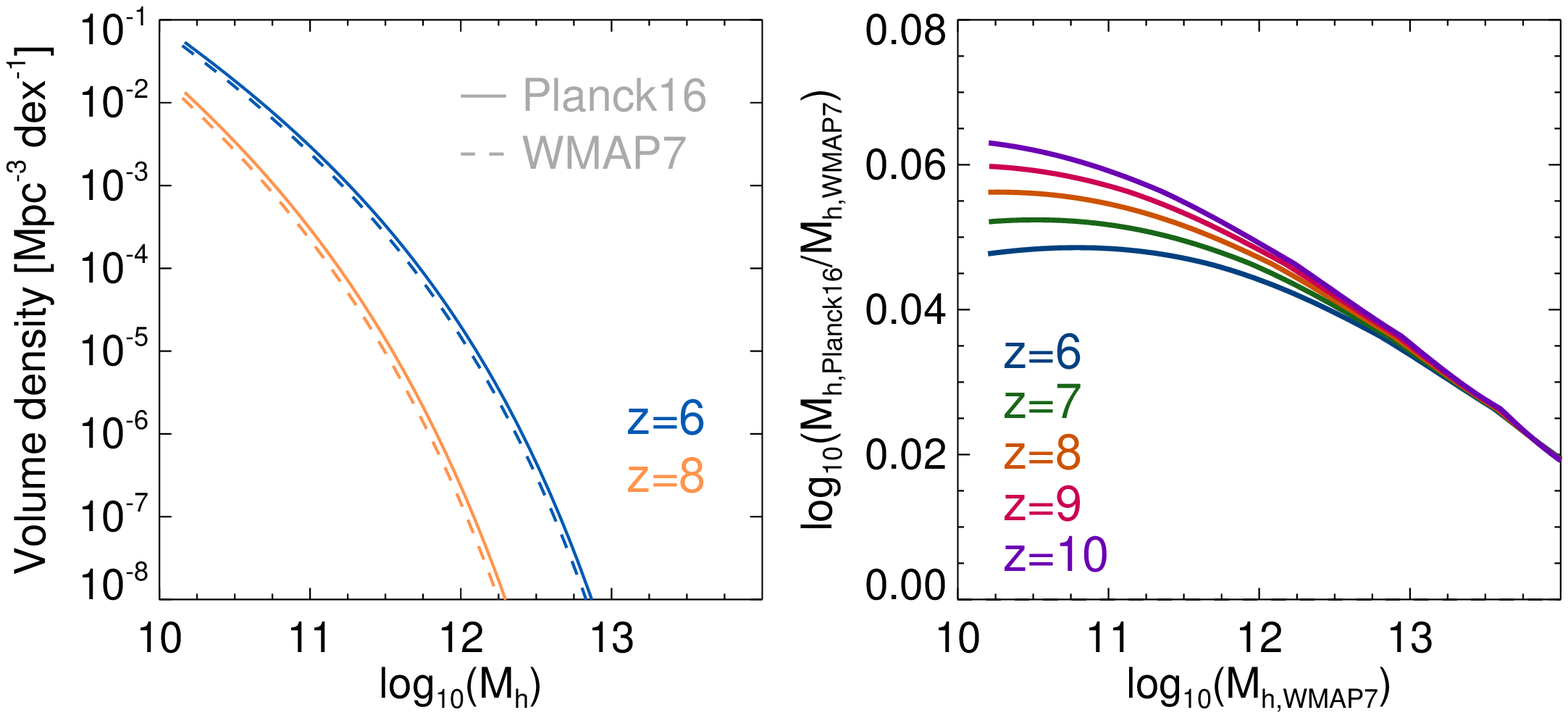}
\caption{		\textbf{ [Left]:} HMF at $z=6$ (blue curves) and $z=8$ (orange) from a \citet{planck2016_cosmology} (solid curves) and WMAP7  \citep{komatsu2011} cosmology.  		\textbf{ [Right]:} Ratio (in log scale) between the halo masses from the WMAP 7 \citep{komatsu2011} cosmology to those from the \citet{planck2016_cosmology} cosmology for the five redshift bins considered in this work, and derived after matching the cumulative number densities of the corresponding halo mass functions. The correction factor depends on the specific redshift but is approximately independent of halo mass within the halo mass range considered in our work ($10^{10.5}$ to $10^{12} \mathcal{M}_\odot$. \label{fig:hmass_cosmology}}
\end{figure*}

In Section \ref{sect:SHMR} we discuss our measurements of the SHMR and compare them to previous observational results and model expectations. Our SHMR estimates and those of \citet{harikane2016, harikane2018,moster2018} and \citet{behroozi2019}  were obtained assuming a cosmology consistent with the \citet{planck2016_cosmology} results, whereas the SHMR estimates of \citet{finkelstein2015b} and \citet{tacchella2018} are based on the WMAP7 cosmology  \citep{komatsu2011}. For illustrative purposes, in the left panel of Figure \ref{fig:hmass_cosmology} we compare the HMF at $z=6$ and $z=8$ from the two cosmologies. Specifically, we adopted the  \citet{behroozi2013} HMF generated by the \textsc{HMFcalc} tool \citep{murray2013}.  When WMAP7 HMFs are adopted for abundance matching procedures, the systematic differences in the volume densities between the two cosmologies translate into $\sim0.05-0.06$ dex lower halo masses and  $\sim15\%$ higher SHMR for WMAP7-based observables compared to Planck-based ones.

To allow for a consistent comparison of all of these estimates, we matched the cumulative volume density of dark matter halos in the two cosmologies and computed the ratio of the corresponding halo masses. The multiplicative factors we used to convert halo masses as presented in \citet{finkelstein2015b} and \citet{tacchella2018} to Planck-based halo masses are presented in the right panel of Figure  \ref{fig:hmass_cosmology}.  These factors depend on redshift but are approximately independent of halo mass within the halo mass range considered in our work ($10^{10.5}$ to $10^{12} \mathcal{M}_\odot$).

\label{sect:app_cosmology}


\bibliographystyle{apj}

\end{document}